\documentclass[12pt]{iopart}
\usepackage{iopams}  
\usepackage{amssymb}
\usepackage{braket}
\usepackage{citesort}
\usepackage{graphicx}
\usepackage{xcolor}
\usepackage{multirow}
\usepackage{booktabs}
\usepackage{cancel}
\newcommand{\vf}{v_\mathrm{F}}
\begin{document}
\title{Spectral properties of interacting helical channels driven by Lorentzian pulses}
\author{Matteo Acciai$^{1,2,3}$, Alessio Calzona$^4$, Matteo Carrega$^{5}$, Thierry Martin$^{3}$ and Maura Sassetti$^{1,2}$}
\address{$^1$ Dipartimento di Fisica, Universit\`a di Genova, Via Dodecaneso 33, 16146, Genova, Italy}
\address{$^2$ CNR-SPIN, Via Dodecaneso 33, 16146, Genova, Italy}
\address{$^3$ Aix Marseille Univ, Universit\'e de Toulon, CNRS, CPT, Marseille, France}
\address{$^4$ Institute for Theoretical Physics and Astrophysics, University of W\"urzburg, 97074 W\"urzburg, Germany}
\address{$^5$ NEST, Istituto Nanoscienze-CNR and Scuola Normale Superiore, Piazza San Silvestro 12, 56127 Pisa, Italy}

\begin{abstract}
Precise shaping of coherent electron sources allows the controlled creation of wavepackets into a one dimensional (1D) quantum conductor. Periodic trains of Lorentzian pulses have been shown to induce minimal excitations without creating additional electron-hole pairs in a single non-interacting 1D electron channel.
The presence of electron-electron (e-e) interactions dramatically affects the non-equilibrium dynamics of a 1D system. Here, we consider the intrinsic spectral properties of a helical liquid, with a pair of counterpropagating interacting channels, in the presence of time-dependent Lorentzian voltage pulses.
We show that peculiar asymmetries in the behavior of the spectral function are induced by interactions, depending on the sign of the injected charges. Moreover, we discuss the robustness of the concept of minimal excitations in the presence of interactions, where the link with excess noise is no more straightforward. Finally, we propose a {scanning} tunneling microscope setup to spectroscopically access and probe the non-equilibrium behavior induced by the voltage drive and e-e interactions. This allows a diagnosis of fractional charges  in a correlated quantum spin Hall liquid in the presence of time-dependent drives.
\end{abstract}

\section{Introduction}
Recent advances in mesoscopic physics allowed to study new quantum effects and their potential for storing, controlling, and manipulating quantum information in a precise and robust fashion~\cite{nayak08,manchon15,wendin2017,haldane2017nobel}. In particular, systems with topological properties, such as the integer and fractional quantum Hall~\cite{klitzing80,girvin1990,stern-review} and topological insulators~\cite{haldane1988prl,bhz06,wu06,konig07qsh,hasan2010colloquium,qi2011topological}, played a prominent role due to  their intrinsic robustness against disorder and decoherence. In these systems, electrons travel ballistically through 1D edge channels~\cite{dolcetto16review}, in the latter case with a definite spin projection linked to the propagation direction, without suffering from backscattering. Topological edge states thus represent an interesting platform to study coherent transport of few electrons, eventually achieving precise control and manipulation of quantum information. This has opened a new field of research, called electron quantum optics (EQO)~\cite{grenier11,bocquillon12eqo,bocquillon14,bauerle2018review}, where the analog of quantum optics setups with single photons and waveguides are now implemented by using fermionic quasiparticles in solid-state quantum conductors. Experimental milestones have already been obtained, with the demonstration of on-demand coherent sources~\cite{feve07,mahe2010,dubois2013levitonsNature} able to inject single electrons in quantum Hall channels~\cite{bocquillon13homscience,bocquillon2013,freulon15hom}. Moreover, interferometric setups with one or more quantum point contacts (QPCs)~\cite{neder06,roulleau07MZ,chalker2007MZ,levkivskyi08modelnu2,kovrizhin2009MZ,parmentier12mesoscopic,wahl14prl,ferraro2014noise,ferraro15antidot,tewari2016,guiducci19,glattli16wavepackets} have also been investigated. Possible extensions to topological insulators, where spin momentum locking plays an important role, have been also theoretically considered~\cite{dolcini11,hofer2013,inhofer2013,ferraro14HOMtopo,calzona15physicaE,dolcetto16entanglement,calzona16energypart,ronetti16,ronetti17polarized,acciai17} and recently realized~\cite{bendias18,strunz19}.

Time-dependent voltage pulses~\cite{dubois2013levitonsNature,dubois13prb} can be exploited to inject single or multiple electrons in a quantum conductor~\cite{grenier13,ferraro13,jullien14tomography,moskalets15,moskalets16,rech16prl,dolcini16chiral,dolcini17,vannucci17heat,ronetti18crystallization,acciai18,ferraro2018review,dolcini18}. However, the underlying presence of a Fermi sea may result in the excitation of electron-hole pairs, producing noisy and unwanted signals~\cite{dubois2013levitonsNature,dubois13prb}. Levitov and coworkers have predicted that by properly engineering the shape of the voltage pulse, it is possible to inject ``clean'' electron wavepackets, whose real-time profile is a Lorentzian pulse, without creating any unwanted electron-hole pair in a non-interacting single-channel quantum conductor~\cite{levitov96,levitov97,keeling06}. Indeed, it has been measured that these voltage pulses, called Levitons, produce zero excess noise, equivalent to the absence of extra electron-hole pairs~\cite{dubois2013levitonsNature,dubois13prb}. {Levitons are currently on the spotlight in EQO and an intense research activity dealing with their peculiar properties is being carried out \cite{battista14,forrester14,forrester15,dasenbrook15,dasenbrook16,moskalets16physe,suzuki17,hofer17,moskalets17-pss,cabart18,moskalets18,dashti19,bisognin2019,burset19}.}

In this paper, we consider a pair of counterpropagating helical channels, subject to a time-dependent drive, where a richer phenomenology arises compared to more conventional single-channel and spinless conductors. This scenario naturally raises the question: how electron-electron interactions affect the dynamics of injected wavepackets and Levitons in helical channels? Are the concepts of clean signal and minimal excitations still valid in the presence of repulsive interactions and counterpropagating channels? In order to answer these questions, we analyze the intrinsic spectral properties of an interacting helical liquid in the presence of a time-dependent drive. We focus on the injection of integer Levitons, the cleanest signal in non-interacting systems, in order to better enlighten {interaction induced effects}. We show that, in this non-equilibrium situation, peculiar charge asymmetries are visible in the spectral function, but they can be masked in a conventional excess noise experiment. Indeed, {although the spectral function and the excess noise are related to each other, they provide totally equivalent information} solely in the absence of interactions.

Spectral properties and fractionalization phenomena of 1D systems can be detected via scanning tunneling spectroscopy \cite{tersoff85,wiesendanger94,chen07}. In particular, {charge partitioning} in Luttinger liquid wires \cite{auslaender02,steinberg07chargefrac,lehur08} has been highlighted, showing that for
an infinite nanotube, fractional charges can be identified through the measurement of both the autocorrelation noise and the cross correlation noise, measured at the extremities of the nanotube~\cite{crepieux03,lebedev05,guigou07,guigou09,guigou09-2,dasprl2011}.
Experimental results of tunneling spectroscopy of topological insulators~\cite{liu15-tip3dti,hus17-tip3dti,voigtlander18-tip,Giordano:18} have also been recently reported.
In this work, we consider a setup involving a polarized tip, whose advantage consists in the possibility to  separately probe all dynamical spectral properties of counterpropagating channels.

The paper is organized as follows: after a pedagogical overview of basic concepts in \Sref{sec:state-of-the-art}, we introduce model and general settings in \Sref{sec:model}. \Sref{sec:spectral} is dedicated to {the analysis of} the out-of-equilibrium spectral functions in the presence of a time-dependent drive. Finally, in \Sref{sec:tip} we propose a possible setup to probe non-equilibrium spectral properties.

\section{Levitons as minimal excitations}\label{sec:state-of-the-art}
For the sake of clarity, it is useful to briefly review and introduce some concepts and definitions regarding time-dependent voltage pulses.
Despite the interest in Lorentzian-shaped pulses is quite recent and has been triggered  by the experimental implementations in EQO~\cite{dubois2013levitonsNature}, the peculiar properties of this kind of drive were theoretically investigated more than two decades ago by Levitov and coworkers~\cite{levitov96,levitov97}. In particular, they answered to the question whether it is possible to generate in a non-interacting 1D quantum conductor a minimal excitation by applying a specific voltage drive $V(t)$. Here, minimality means that the generation of electron- or of hole-like excitations by the drive produces no additional neutral quasiparticle, i.e. electron-hole pairs. In other words, the number of extra holes or electrons generated by the drive must vanish. Levitov's result~\cite{levitov96,levitov97} states that a minimal Lorentzian pulse carrying an integer particle number $q=(-e/h)\int \rmd t V(t)$ (or a superposition of such pulses with integer charges of the same sign) produces minimal excitations. In particular, we will focus on a periodic train (from here on we set $\hbar=1$)
\begin{equation}
V(t)=\sum_{j\in\mathbb{Z}}V^{(0)}(t-jT)=\sum_{j\in\mathbb{Z}}\frac{q}{(-e)}\frac{2w}{w^2+(t-jT)^2}\,,
\label{eq:periodic}
\end{equation}
with $T$ the period, $w$ the width of each pulse and $q$ an integer number representing the charge per period (in units of $-e$) of the drive.

Here, we consider a 1D non-interacting system with linear dispersion $\epsilon=\vf k$, $\vf$ being the Fermi velocity. Let us focus in this Section on the case of a minimal electronic excitation by choosing $q>0$. We also consider the zero temperature case, in order to get rid of possible thermally-excited electron-hole pairs. The requirement that no extra holes are generated can be written as $N_h=\sum_{k<k_\mathrm{F}}\braket{\hat C_k\hat C^\dagger_k}=0$, where $k_\mathrm{F}$ is the Fermi momentum and $\hat C_k$ is the annihilation operator for an electron with momentum $k$. {Equivalently, }the variation in the electron occupation number (or electron momentum distribution) $\Delta n(k)=\braket{\hat C^\dagger_k \hat C_k}-\braket{\hat c^\dagger_k \hat c_k}$ must be positive for every $k$. Here, $\hat c_k$ is the annihilation operator in the equilibrium situation, when no drive is present. Thus, by variation (indicated with the notation $\Delta$) we mean that we consider just the effects induced by the drive, by properly subtracting the equilibrium contribution~\cite{grenier11,ferraro13,bocquillon14,rech16prl,dolcini16chiral,dolcini17,ronetti18crystallization}. Furthermore, since the system is non-interacting, an equivalent information is provided by the variation of the lesser local spectral function, $\Delta\mathcal{A}^<(\omega;x)$.
The latter is defined as
\begin{equation}
\mathcal{A}^<(\omega;x)=-\rmi\int_{-T/2}^{T/2}\frac{\rmd t}{T}\int_{-\infty}^{+\infty}\frac{\rmd\tau}{2\pi}\,\rme^{\rmi\omega\tau} G^<\left(t+\frac{\tau}{2},t-\frac{\tau}{2};x\right)\,,
\label{eq:spectral-def}
\end{equation}
where the corresponding Green function is
\begin{equation}
G^<(t_1,t_2;x)=\rmi\braket{\hat{\Psi}^\dagger(x,t_2)\hat{\Psi}(x,t_1)}\,,
\end{equation}
with $\hat\Psi(x,t)$ denoting the time evolution of the fermionic field in the Heisenberg picture. {Due to non-stationary effects induced by $V(t)$, the Green function in \eref{eq:spectral-def} depends also on $t$ and not only on $\tau$. It is then a standard procedure to introduce the time average over one period of the drive \cite{grenier11,grenier11tomography}. This is also relevant from an experimental point of view \cite{bocquillon14}, as single-shot measurements are currently out of reach and averaging over several driving periods is necessary.} The variation of the spectral function is $\Delta\mathcal{A}^<(\omega;x)=\mathcal{A}^<(\omega;x)-\mathcal{A}_0^<(\omega;x)$, with $\mathcal{A}_0^<(\omega;x)$ the equilibrium spectral function, defined as in \eref{eq:spectral-def} by replacing the Green function with the equilibrium one $G_0^<(t_1,t_2;x)=\rmi\braket{\hat\psi^\dagger(x,t_2)\hat\psi(x,t_1)}$. Here, $\hat\psi(x,t)$ is the time evolution of the fermionic field at equilibrium, with no applied drive.

In a 1D non-interacting system, the dispersion relation directly connects $\Delta n(k)$ and $\Delta\mathcal{A}^<(\omega;x)$, which therefore provide equivalent information, so that a minimal excitation is associated with a positive-definite excess lesser spectral function $\Delta\mathcal{A}^<(\omega;x)$.
In addition, from \Eref{eq:spectral-def}, it is possible to {demonstrate} the following sum rule:
\begin{equation}
\int_{-\infty}^{+\infty}\rmd\omega\,\Delta\mathcal{A}^<(\omega;x)=\frac{q}{\vf T}\,.
\label{eq:sum1}
\end{equation}
In summary, while the integral of $\Delta\mathcal{A}^<(\omega;x)$ is always the same, only a positive-definite $\Delta\mathcal{A}^<(\omega;x)$ is a signature of a minimal excitation and regions where this spectral function is negative are related to unwanted particle-hole pairs.

\begin{figure}[htbp]
	\begin{center}
		\includegraphics[width=0.7\textwidth]{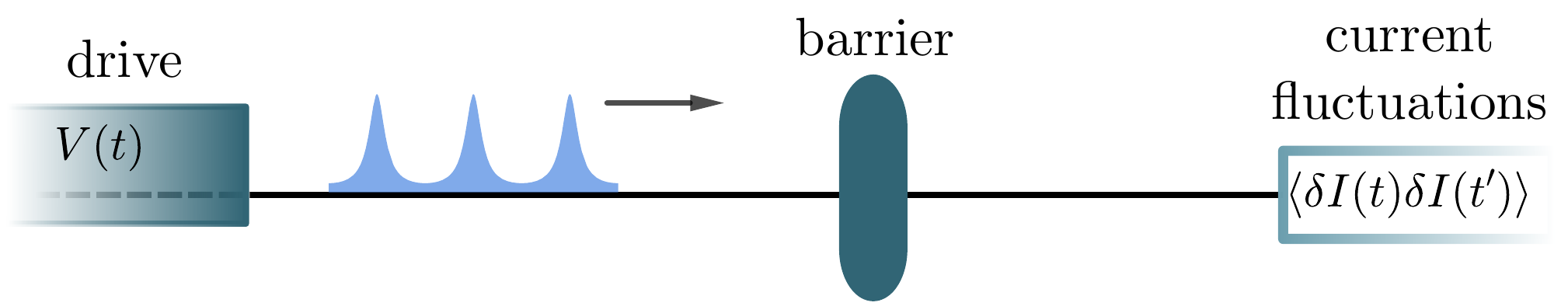}
		\caption{Excitations generated in a 1D non-interacting ballistic conductor by a periodic drive $V(t)$ are partitioned by a tunnel barrier (for instance a quantum point contact). Current fluctuations due to this tunneling process are measured after the barrier and can be directly linked to the concept of ``minimal excitation''.}
		\label{fig:sketch}
	\end{center}
\end{figure}
From an experimental point of view, signatures of this minimality can be found by considering the zero-frequency noise $S$ produced when the generated excitations are partitioned by a tunnel barrier (typically a QPC~\cite{dubois2013levitonsNature,glattli16wavepackets}), as sketched in \Fref{fig:sketch}. It is defined as
\begin{equation}
S=2\int_{-T/2}^{T/2}\frac{\rmd t}{T}\int_{-\infty}^{+\infty}\rmd t'\braket{\delta \hat I(t)\delta \hat I(t+t')}\,,
\end{equation}
where $\delta \hat I(t)=\hat I(t)-\braket{\hat I(t)}$ denotes fluctuations with respect to the average current. Indeed, it was shown~\cite{keeling06,dubois13prb} that at zero temperature the number of extra holes excited by the drive is proportional to the excess noise
\begin{equation}
S_\mathrm{exc}=S-2eI=S-2e\int_{-T/2}^{T/2}\frac{\rmd t}{T}\braket{\hat I(t)}\,.
\end{equation}
Therefore, the excess noise vanishes if the drive $V(t)$ generates minimal electron excitations. This feature was experimentally observed~\cite{dubois2013levitonsNature}.

As far as interacting systems are concerned, the situation becomes more complicated. First of all, $\Delta n(k)$ and $\Delta\mathcal{A}^<(\omega ; x)$ are no more related to each other; still, the sum rule in \Eref{eq:sum1} has to be satisfied. Therefore, it is natural to refer to an excitation as minimal if its spectral function has a definite sign. 
As we will discuss, this notion of minimality is in general unrelated to a vanishing excess noise, which is instead due to a different property of the spectral function. To fully characterize the behavior of Levitons in an interacting system we therefore will focus on its intrinsic spectral properties out of equilibrium.

\section{Model and general setting}
\label{sec:model}
We consider a pair of interacting helical channels, capacitively coupled to an external voltage source. The Hamiltonian is $\hat H=\hat H_\mathrm{HLL}+\hat H_g$.
The Hamiltonian of the helical channels reads $\hat{H}_{\mathrm{HLL}}=\hat{H}_0+\hat{H}_\mathrm{int}$, where
\begin{equation}
\hat H_0=\sum_{r=R,L}\int_{-\infty}^{+\infty} \rmd x\,\hat \Psi_r^\dagger(x)(-\rmi\vartheta_r\vf\partial_x)\hat \Psi_r(x)\,,\quad \vartheta_{R/L}=\pm 1\,,
\end{equation}
describes a pair of free counterpropagating channels, with right- $(\hat{\Psi}_R)$ and left- $(\hat{\Psi}_L)$ moving electrons having spin up and down, respectively, due to the so-called spin-momentum locking~\cite{qi2011topological,ronetti16}.
The interacting part $\hat{H}_\mathrm{int}$ accounts for short range Coulomb interactions~\cite{voit1995}:
\begin{equation}
\hat H_\mathrm{int}=\frac{g_4}{2}\sum_{r=R,L}\int_{-\infty}^{+\infty} \rmd x\,[\hat n_r(x)]^2+g_2\int_{-\infty}^{+\infty} \rmd x\, \hat n_R(x)\hat n_L(x)\,,
\end{equation}
with $g_2$ and $g_4$ the inter- and intra-channel coupling constants, respectively, and $\hat n_r=\,:\!\!\hat\Psi_r^\dagger\hat\Psi_r\!\!:$ the particle density operators. {Within the Luttinger liquid theory, the interaction strength is described by the Luttinger parameter
\begin{equation}
K=\sqrt{\frac{2\pi\vf+g_4-g_2}{2\pi\vf+g_4+g_2}}\,,
\label{eq:K}
\end{equation}
which is a function of $g_2$ and $g_4$\footnote{For a Galilean invariat system one has $g_2=g_4=g$ and then the coupling constant can be estimated from the knowledge of $K$ and $\vf$.}. In the non-interacting case $K=1$, while repulsive interactions $K<1$.
Recent experimental results suggest that interactions can be relevant in the quantum spin Hall (QSH) state: a Luttinger parameter $K=0.42$ was reported in Bismuthene on SiC substrate \cite{stuhler19}. In addition, a previous work also reported the evidence for interaction effects in the QSH state realized in an InAs/GaSb quantum well. Although the value of $K$ in that system is debated -- $K=0.22$ \cite{li2015helical} vs.\ $K=0.8$ \cite{glazman16} -- interactions seem to be anyway relevant.}

It is useful to introduce bosonic operators $\hat\Phi_r(x)$ via the bosonization identity~\cite{vondelft,giamarchi}
\begin{equation}
\hat\Psi_r(x)=\frac{\hat F_r}{\sqrt{2\pi a}}\,\rme^{\rmi\vartheta_r k_\mathrm{F}x}\rme^{-\rmi\sqrt{2\pi}\hat\Phi_r(x)}\,,
\label{eq:bos}
\end{equation}
with $\hat F_r$ a Klein factor ensuring the proper fermionic anticommutation rules and $a$ a short length cutoff. The  Hamiltonian $\hat{H}_\mathrm{HLL}$ becomes
\begin{equation}
\hat H_\mathrm{HLL}=\frac{u}{2}\sum_{\eta=\pm}\int_{-\infty}^{+\infty} \rmd x\,[\partial_x\hat\Phi_\eta(x)]^2\,,
\end{equation}
where $u=(2\pi)^{-1}\sqrt{(2\pi\vf+g_4)^2+g_2^2}$ represents the renormalized velocity and $\hat\Phi_\eta$ are new chiral bosonic fields, related to $\hat\Phi_r$ via the relation
\begin{equation}
\hat{\Phi}_{r}(x)=\sum_{\eta=\pm}A_{\eta\vartheta_r}\hat{\Phi}_{\eta}(x)\,,\qquad A_\pm=\frac{1}{2}\left(\frac{1}{\sqrt{K}}\pm\sqrt{K}\right)\,.
\label{eq:phi-eta}
\end{equation}
For $K=1$ we have $\hat\Phi_\pm=\hat\Phi_{R/L}$.

The helical channels are capacitively coupled to an external gate
\begin{equation}
\hat H_g=-e\int_{-\infty}^{+\infty} \rmd x\,U(x,t)[\hat n_R(x)+\hat n_L(x)]\,,
\end{equation}
where $U(x,t)=F(x)V(t)$ encodes the spatial and temporal profile of the external voltage. Hereafter we consider an extended gate, by choosing $F(x)=\Theta(-x)$ (with $\Theta(x)$ the Heaviside step function), so that the drive $V(t)$ is applied in the region $(-\infty,0)$.
In the following we will consider a periodic train of pulses with period $T$, as in \Eref{eq:periodic}, specifying only later the precise form of the pulse $V^{(0)}(t)$. 
{The} charge per period of the pulse (in units of $-e$) is
\begin{equation}
q=\int_{-T/2}^{T/2}\frac{-e}{2\pi}V(t)\,\rmd t\,.
\label{eq:q-def}
\end{equation}

The equations of motion for fields $\hat\Phi_\eta$, obtained from the full Hamiltonian $\hat H$, are
\begin{equation}
(\partial_t+u\eta\partial_x)\hat\Phi_\eta(x,t)=-e\sqrt{\frac{K}{2\pi}}\Theta(-x)V(t)
\end{equation}
whose solution can be written as
\numparts
\begin{eqnarray}
\fl\hat\Phi_+(x,t)&=\hat\phi_+(x-ut,0)-e\sqrt{\frac{K}{2\pi}}\left[\Theta(x)\int_{-\infty}^{t-\frac{x}{u}}\rmd t'V(t')+\Theta(-x)\int_{-\infty}^{t}\rmd t'V(t')\right]\\
\fl\hat\Phi_-(x,t)&=\hat\phi_-(x+ut,0)-e\sqrt{\frac{K}{2\pi}}\Theta(-x)\int_{t+\frac{x}{u}}^t\rmd t'V(t')
\end{eqnarray}
\label{eq:eom-sol}
\endnumparts
with $\hat\phi_\pm(x\mp ut,0)$ the chiral evolution of bosonic fields without external drive.
The time evolution of fermion operators $\hat\Psi_r(x,t)$ is thus obtained by using \eref{eq:phi-eta} and \eref{eq:bos}. Finally, a generic expectation value of an operator $\hat O(t)$, is obtained as $\braket{\hat O(t)}=\Tr[\hat{\varrho} \hat O(t)]$. Here, $\hat{\varrho}$ is the time-independent equilibrium density matrix at $t=-\infty$ when no voltage is applied and thus originating only from the Hamiltonian $\hat H_\mathrm{HLL}$.

\subsection{Excess particle density}
\begin{figure}[hbtp]
	\begin{center}
		\includegraphics[width=0.8\textwidth]{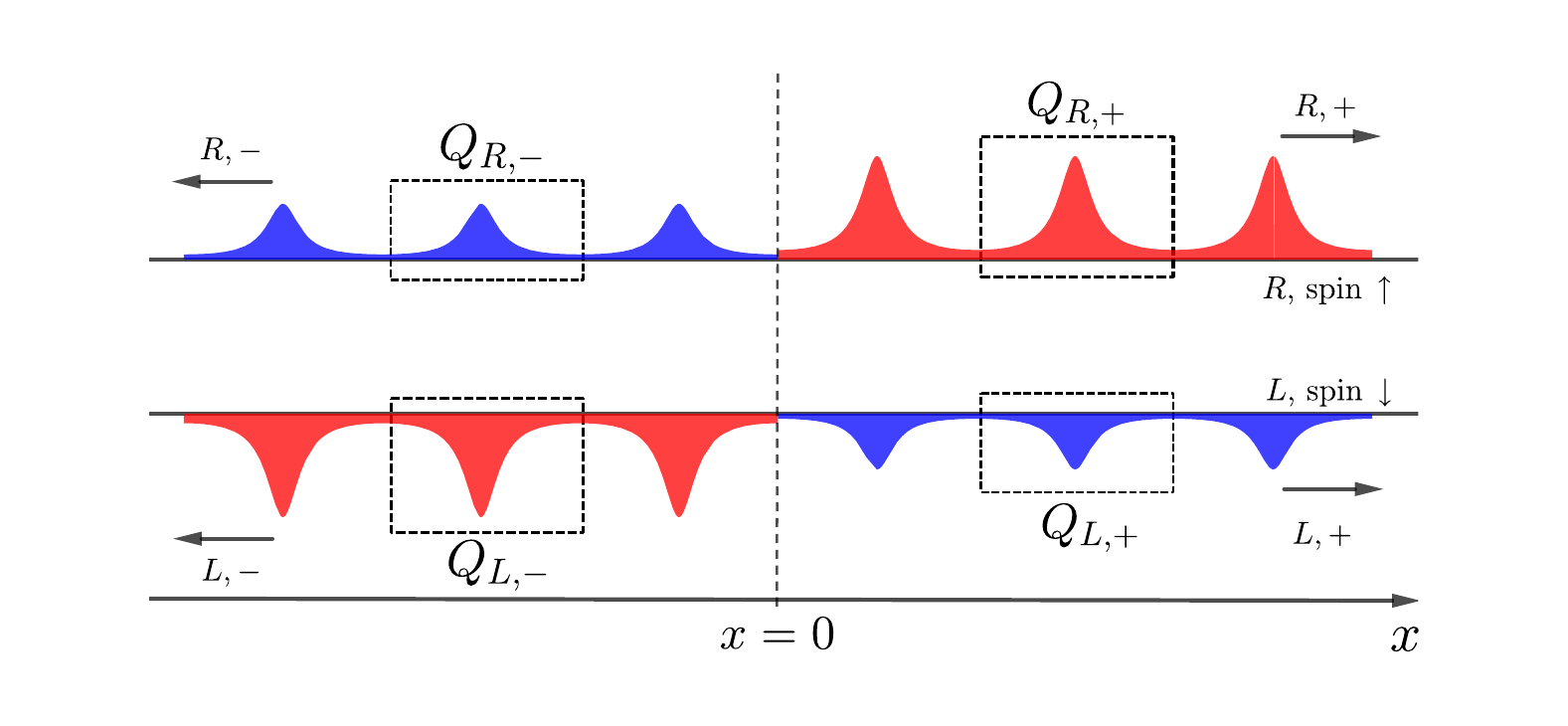}
		\caption{Sketch of the space profile of the excess charge density due to the periodic voltage $V(t)$, with $q>0$. Excitations originating at $x=0$ on both channels $r=R,L$ propagate in the positive or negative direction of the $x$ axis, depending on the value of the index $\eta$. The charge per period (in units of $-e$) carried by each excitation is $Q_{r,\eta}$, as given in \Eref{eq:charge}. Red pulses indicate excitations which are also present in the non-interacting case, while blue ones refer to those originating only as a result of interactions.}
		\label{fig:packets}
	\end{center}
\end{figure}

In order to understand how excitations are generated {by} the drive, a first quantity to look at is the average particle density $\Braket{\hat n(x,t)}=\Braket{\hat n_R(x,t)+\hat n_L(x,t)}$.
In particular, we are interested in the deviations from the equilibrium situation
\begin{equation}
\Delta n(x,t)=\Braket{\hat n(x,t)}-\Braket{\hat n_0(x,t)}\,,
\end{equation}
with $\hat n_0$ denoting the particle density operator in the absence of the drive. It is useful to introduce the chiral particle density operators $\hat n_\eta(x,t)=-\eta\sqrt{K/(2\pi)}\partial_x\hat\Phi_\eta(x,t)$, since the total particle density can be written as $\hat n(x,t)=\hat n_+(x,t)+\hat n_-(x,t)$. From \Eref{eq:eom-sol} we find
\begin{equation}
\Delta n(x,t)=\sum_{\eta=\pm}\Delta n_\eta(x,t)=-e\sum_{\eta=\pm}\frac{\eta K}{2\pi u}V\left(t-\eta\frac{x}{u}\right)\Theta(\eta x)\,.
\label{eq:density-eta}
\end{equation}
This shows that the effect of the drive is to induce excitations propagating both to the right $(\eta=+)$ and to the left $(\eta=-)$. Notice that this expression is independent of the temperature and that the chiral right- (left-) moving excitation contributes only at $x>0$ $(x<0)$, see \Fref{fig:packets}.
Moreover, from the relation
\begin{equation}
\Delta n_r(x,t)=\frac{\vartheta_r}{K}\left[\frac{1+\vartheta_r K}{2}\Delta n_+(x,t)-\frac{1-\vartheta_r K}{2}\Delta n_-(x,t)\right]\,,
\label{eq:density-r-eta}
\end{equation}
excitations for each chirality $\eta$ are composed of contributions coming from both channels $r=R,L$.
By combining \eref{eq:density-eta} and \eref{eq:density-r-eta}, the spatial profile of the excitation on channel $r$ moving in the $\eta$ direction is
\begin{equation}
\Delta n_{r,\eta}(x,t)=\frac{-e\vartheta_r}{2\pi u}\frac{1+\eta\vartheta_r K}{2}V\left(t-\eta\frac{x}{u}\right)\Theta(\eta x)\,,
\end{equation}
as schematically depicted in \Fref{fig:packets}.
It is worth {to underline }that all these contributions can be distinguished, since $R$ and $L$ channels have opposite spin projections.

In the non-interacting case $K=1$, only $\Delta n_{R,+}$ and $\Delta n_{L,-}$ are present (red pulses), consistently with the fact that for free fermions $R$ and $L$ channels are right- and left-moving, respectively, and do not mix. On the contrary, in the interacting case also the $(R,-)$ and $(L,+)$ channels are involved (blue pulses), due to charge fractionalization phenomena~\cite{safi95,pham2000,auslaender02,auslaender05,jompol09,deshpande10,perfetto14timeresolved}. {We underline that here, following the Luttinger liquid literature, the word fractionalization is used as a synonym of partitioning. Therefore, it does not refer to the presence of stable fractionally-charged quasiparticle excitations, as the ones which instead characterize the fractional QHE.}
The charge\footnote{With a little abuse of language, we shall systematically refer to $Q_{r,\eta}$ as the charge of the excitations, actually meaning the charge in units of $-e$.} per period $Q_{r,\eta}$ carried on each channel is obtained by integrating over one period the corresponding contribution to the current flowing away from the point $x=0$. Thus, by fixing a detection point $d>0$, we find
\begin{equation}
Q_{r,\eta}=u\int_{-T/2}^{T/2}\Delta n_{r,\eta}(\eta d,t)\,\rmd t=\vartheta_r\frac{1+\eta\vartheta_r K}{2}\,q=\vartheta_r\sqrt{K}A_{\eta\vartheta_r}q\,,
\label{eq:charge}
\end{equation}
where \Eref{eq:q-def} has been used and coefficients $A_{\eta\vartheta_r}$ are defined in \eref{eq:phi-eta}. The charge $Q_{r,\eta}$ carried by each excitation is an interaction-dependent fraction of the charge $q$ injected by the drive $V(t)$, which is the experimentally tunable parameter. Notice that $\pm q$ would be the charge carried in the non-interacting system $(K=1)$ by the excitation on channel $(R,+)/(L,-)$. We emphasize that the effect of interactions goes well beyond the simple renormalization of the charge carried on each channel. Indeed, interacting correlation functions contain interaction-dependent power-laws which are not present in the K=1 case \cite{voit1995,giamarchi}. As detailed in the next Section, this leads to qualitative differences in the spectral properties between interacting and non-interacting systems.
\section{Non-equilibrium spectral function}
\label{sec:spectral}
\subsection{General properties}
In this Section we discuss the intrinsic spectral properties~\cite{vondelft,voit1995,guinea95,calzona17quench,calzona17-2} of the interacting system in the presence of a periodic drive. To this end, we will focus on the variation of the spectral function, with respect to its equilibrium value (when $V(t)=0$). It is therefore useful to introduce the variation of the lesser/greater local Green function:
\numparts
\begin{eqnarray}
\fl-\rmi\Delta G_r^<(t_1,t_2;x)&=\Braket{\hat\Psi_r^\dagger(x,t_2)\hat\Psi_r(x,t_1)}\rme^{\rmi e\int_{t_1}^{t_2}U(x,t')\rmd t'}-\Braket{\hat\psi_r^\dagger(x,t_2)\hat\psi_r(x,t_1)}\,,\\
\fl+\rmi\Delta G_r^>(t_1,t_2;x)&=\Braket{\hat\Psi_r(x,t_1)\hat\Psi_r^\dagger(x,t_2)}\rme^{\rmi e\int_{t_1}^{t_2}U(x,t')\rmd t'}-\Braket{\hat\psi_r(x,t_1)\hat\psi_r^\dagger(x,t_2)}\,.
\end{eqnarray}
\endnumparts
Here, $\hat\psi_r(x,t)$ denotes the time evolution of the fermion operator for $r$-electrons in the absence of the drive. The exponential factor is a Wilson line, ensuring gauge invariance of the correlators~\cite{dolcini16chiral,dolcini18}.
{Due to $V(t)$}, the Green functions depend both on the difference $\tau=t_1-t_2$ and on the average time $t=(t_1+t_2)/2$. Therefore, we define the local (excess) spectral functions as the Fourier transform with respect to $\tau$ and we further average over the period of the drive $T$:
\begin{equation}
\Delta\mathcal{A}_r^\gtrless(\omega;x)=\pm\rmi\int_{-T/2}^{T/2}\frac{\rmd t}{T}\int_{-\infty}^{+\infty}\frac{\rmd\tau}{2\pi}\,\rme^{\rmi\omega\tau}\Delta G_r^\gtrless\left(t+\frac{\tau}{2},t-\frac{\tau}{2};x\right)\,.
\label{eq:def-spectral}
\end{equation}

By resorting to standard bosonization techniques~\cite{vondelft,giamarchi,gambetta:epl}, Green functions can be written as
\begin{equation}
\Delta G_r^\gtrless\left(t+\frac{\tau}{2},t-\frac{\tau}{2};x\right)=\sum_{\eta=\pm}\Theta(\eta x)\Delta G_{r,\eta}^\gtrless\left(t+\frac{\tau}{2},t-\frac{\tau}{2}\right)\,,
\label{eq:spectral-time-r}
\end{equation}
where the contribution of the excitation on the channel $(r,\eta)$ has the structure
\begin{equation}
\Delta G_{r,\eta}^\gtrless\left(t+\frac{\tau}{2},t-\frac{\tau}{2}\right)=\pm\rmi G_0^\gtrless(\tau)P^\gtrless_{r,\eta}(\tau)\Xi_{r,\eta}(t,\tau)\,.
\label{eq:spectral-time-r-eta}
\end{equation}
Due to the function $\Theta(\eta x)$, the term related to the excitation with $\eta=+\,(-)$ contributes only at positive (negative) values of $x$. Taking advantage of this fact, we can write $\Delta\mathcal{A}_{r}^\gtrless(\omega;x)=\sum_{\eta=\pm}\Theta(\eta x)\Delta\mathcal{A}_{r,\eta}^\gtrless(\omega)$, defining the greater/lesser spectral function $\Delta\mathcal{A}^\gtrless_{r,\eta}(\omega)$ associated with the excitation $(r,\eta)$ as
\begin{equation}
\Delta\mathcal{A}_{r,\eta}^\gtrless(\omega)=\pm\rmi\int_{-T/2}^{T/2}\frac{\rmd t}{T}\int_{-\infty}^{+\infty}\frac{\rmd\tau}{2\pi}\rme^{\rmi\omega\tau}G_0^\gtrless(\tau)P^\gtrless_{r,\eta}(\tau)\Xi_{r,\eta}(t,\tau).
\end{equation}
Here,
\begin{equation}
G_0^<(\tau)=\rmi\Braket{\hat\psi_r^\dagger(0)\hat\psi_r(\tau)}=\frac{\rmi}{2\pi(a-\rmi u\tau)}\left[\frac{a}{a-\rmi u\tau}\right]^{2A_-^2}=-G_0^>(-\tau)
\label{eq:corr-eq}
\end{equation}
represents the equilibrium Green function at zero temperature and is independent of the channel index $r$.
The term $P_{r,\eta}^<(\tau)=(\rmi\eta u\tau-a\vartheta_r)(\rmi\eta u\tau)^{-1}=P_{r,\eta}^>(-\tau)$ stems from the point splitting procedure and ensures that the diagonal limit $\tau\to 0$ correctly reproduces the excess particle density $\Delta n_r(x,t)=\lim_{\tau\to 0}\Delta G_r^<(t+\tau/2,t-\tau/2;x)$, as already discussed in~\cite{acciai17}.
This factor is only relevant at small values of $\tau$ and thus affects the corresponding spectral function $\Delta\mathcal{A}^\gtrless_{r,\eta}(\omega)$ at high energies.
The effects of the time-dependent external drive are encoded in the phase factor
\begin{equation}
\Xi_{r,\eta}(t,\tau)=\exp\left[-\rmi e\eta\sqrt{K}A_{\eta\vartheta_r}\int_{t+\tau/2}^{t-\tau/2}V(t')\,\rmd t'\right]-1\,,
\label{eq:phase}
\end{equation}
where we recognize the {partitioning or} fractionalization factors $\sqrt{K}A_{\eta\vartheta_r}$ already encountered when discussing the excess particle density, see also \eref{eq:phi-eta}.
In the presence of a periodic drive, the phase factor can be conveniently decomposed in a Fourier series~\cite{dubois13prb,ferraro18squeezing}
\begin{equation}
\rme^{\rmi e\int_{-\infty}^{t}V(t')\,\rmd t'}=\rme^{-\rmi q\Omega t}\sum_{\ell\in\mathbb{Z}}p_\ell(q)\,\rme^{-\rmi\ell\Omega t}\,.
\label{eq:Fourier}
\end{equation}
Here, the photoassisted amplitudes $p_\ell$ represent the probability amplitude for an electron to emit $(\ell<0)$ or absorb $(\ell>0)$ $|\ell|$ photons of energy $\Omega$ as a consequence of the ac drive. Their particular expression depends on the functional form of $V(t)$.
By using \eref{eq:Fourier} it is easy to show that
\begin{equation}
\int_{-T/2}^{T/2}\frac{\rmd t}{T}\,\Xi_{r,\eta}(t,\tau)=\sum_{\ell\in\mathbb{Z}}|p_\ell(Q_{r,\eta})|^2\left(\rme^{-\rmi\eta\vartheta_r\tau(\ell+Q_{r,\eta})\Omega}-1\right)\,,
\label{eq:xi-integral}
\end{equation}
so that $\Delta\mathcal{A}_{r,\eta}^<(\omega)$ reads
\begin{equation}
\fl\Delta\mathcal{A}_{r,\eta}^<(\omega)=\frac{a^{2A_-^2}}{4\pi^2}\sum_{\ell\in\mathbb{Z}}|p_\ell(Q_{r,\eta})|^2\int_{-\infty}^{+\infty}\rmd\tau\frac{\rme^{\rmi\tau(\omega-\eta\vartheta_r\Omega_\ell)}-\rme^{\rmi\omega\tau}}{(a-\rmi u\tau)^{1+2A_-^2}}\frac{\rmi\eta u\tau-a\vartheta_r}{\rmi\eta u\tau}\,,
\label{eq:lesser-spectral}
\end{equation}
with $\Omega_\ell=(\ell+Q_{r,\eta})\Omega$. Notice that the photoassisted coefficients depend on the charge $Q_{r,\eta}$ of the different excitations.

Now, some general properties can be discussed. First of all, the following sum rules are fulfilled:
\begin{equation}
Q_{r,\eta}=uT\int_{-\infty}^{+\infty}\Delta\mathcal{A}_{r,\eta}^<(\omega)\,\rmd\omega=-uT\int_{-\infty}^{+\infty}\Delta\mathcal{A}_{r,\eta}^>(\omega)\,\rmd\omega\,.
\label{eq:sum-rule}
\end{equation}
This indicates that the integral over energies of the lesser excess spectral function on channel $r,\eta$ gives the charge per period carried by the excitation on that channel, in the same way as the integral over time of the excess charge density [see \Eref{eq:charge}].
{Due to \Eref{eq:sum-rule}, we can introduce the notion of minimality of an excitation, by requiring that the corresponding excess spectral function has everywhere the same sign as the one dictated by its sum rule.
Physically, every $\Delta\mathcal{A}^{\gtrless}_{r,\eta}(\omega)$ represents a perturbation (with respect to equilibrium) which is globally larger if somewhere the function has a different sign compared to what \Eref{eq:sum-rule} requires, showing the presence of additional positive/negative charges.}

Moreover, looking at Equations \eref{eq:corr-eq} and \eref{eq:phase} some symmetry relations between greater and lesser components of spectral functions can be deduced. In particular, we have
\begin{equation}
\Delta\mathcal{A}_{r,\eta}^<(\omega,q)=\Delta\mathcal{A}_{-r,-\eta}^<(\omega,-q)=\Delta\mathcal{A}_{r,\eta}^>(-\omega,-q)\,.
\label{eq:relations-spectral}
\end{equation}
Here, for the sake of clarity, we have explicitly included the dependence on the parameter $q$ [see \eref{eq:q-def}]. We also used the notation $-R=L$ and viceversa. Going from positive to negative $q$ simply means flipping the sign of $V(t)$. Thanks to \Eref{eq:relations-spectral} we can simply focus on $\Delta\mathcal{A}_{R,\pm}^<(\omega,q)$ and obtain from these all other contributions by properly changing the sign of $q$ and $\omega$. Finally, in the non-interacting case $K=1$, some additional symmetry relations are satisfied:
\begin{equation}
\fl\Delta\mathcal{A}_{r,\eta}^<(\omega,q)=-\Delta\mathcal{A}_{r,\eta}^>(\omega,q)=-\Delta\mathcal{A}_{r,\eta}^<(-\omega,-q)\,,\qquad (r,\eta)=(R,+)\,\,\mathrm{or}\,\,(L,-)\,.
\label{eq:symmetry-k1}
\end{equation}
As a consequence, the total excess spectral function, defined as
\begin{equation}
\fl\Delta\mathcal{A}_{r,\eta}(\omega,q)=\Delta\mathcal{A}_{r,\eta}^<(\omega,q)+\Delta\mathcal{A}_{r,\eta}^>(\omega,q)=\Delta\mathcal{A}_{r,\eta}^<(\omega,q)+\Delta\mathcal{A}_{r,\eta}^<(-\omega,-q)\,,
\label{eq:total-spectral}
\end{equation}
vanishes when $K=1$ independently of the drive. 
In the presence of interactions, instead, $\Delta\mathcal{A}_{r,\eta}(\omega)\ne 0$.
Therefore, a measure sensitive to $\Delta\mathcal{A}_{r,\eta}(\omega)$ would be able to clearly distinguish between an interacting and a non-interacting system, see \Sref{sec:tip}.

\subsection{Lorentzian pulses}
As summarized in \Sref{sec:state-of-the-art}, periodic trains of Lorentzian pulses play a special role in the context of EQO~\cite{dubois2013levitonsNature,glattli16wavepackets} (being the best choice for the generation of minimal excitations for non-interacting systems) and have triggered an intense research activity~\cite{dubois13prb,ferraro14decoherence,moskalets16,rech16prl,ronetti18crystallization,ferraro18squeezing,acciai19,dasenbrook15,dasenbrook16,moskalets16physe,hofer17,moskalets17-pss,cabart18,moskalets18,dashti19,bisognin2019,burset19}.
Here, we will thus focus on this particular shape of wavepackets, in order to elucidate the effects caused by the presence of repulsive interactions.
The real-time shape of a single Lorentzian pulse is given by
\begin{equation}
V^{(0)}(t)=\frac{q}{-e}\frac{2w}{w^2+t^2}\,.
\label{eq:single-pulse}
\end{equation}
In order to calculate the spectral functions we need to specify the photoassisted coefficients appearing in \eref{eq:lesser-spectral} and then compute the integral.
They read~\cite{dubois13prb,rech16prl}
\begin{equation}
p_\ell(q)=q\sum_{s=0}^{+\infty}\frac{{(-1)}^s\,\Gamma(q+\ell+s)\,\rme^{-2\pi w(2s+\ell)/T}}{\Gamma(q+1-s)\Gamma(1+s)\Gamma(1+\ell+s)}\,.
\label{eq:pl}
\end{equation}

We now restrict our attention to the case of Lorentzian pulses with integer values of $Q_{r,\eta}$ for the corresponding channel $(r,\eta)$.
As a consequence of the particular form of the photoassisted coefficients, the spectral function shows some remarkable properties for these integer values.
\begin{figure}[htbp]
	\begin{center}
		\includegraphics[width=\textwidth]{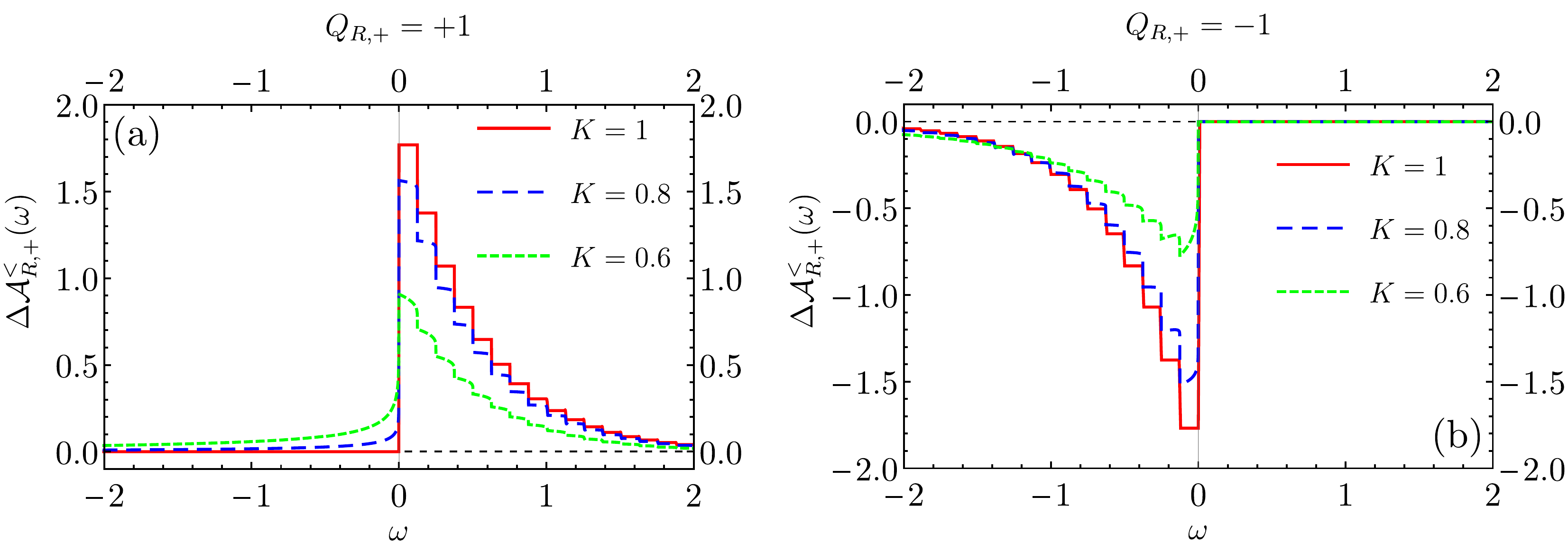}
	\end{center}
	\caption{Lesser spectral function $\Delta\mathcal{A}_{R,+}^<(\omega)$, in units of $w(uT)^{-1}$, as a function of $\omega$, in units of $w^{-1}$. Charges are $Q_{R,+}=\pm 1$, as specified on each panel and different values of the interaction strength $K$ are indicated in the label. In all plots we set a representative value for the period $T=50 w$ and $a=0.01uw$.}
	\label{fig:plots-lesser-Rplus}
\end{figure}
The behavior of the lesser spectral function $\Delta \mathcal{A}_{R,+}^<$ for right-moving wavepackets is shown in \Fref{fig:plots-lesser-Rplus} for different values of the interaction strength $K$. Note that this contribution is present also in the non-interacting case ($K=1$), where $Q_{R,+}=q$. Different panels correspond to integer but opposite value of the injected charge $Q_{R,+}=\pm1$.
In the absence of interactions ($K=1$), {\Eref{eq:symmetry-k1} dictates that the spectral function with $Q_{R,+}=-1$ can be obtained }{by  reversing the one with $Q_{R,+}=+1$ with respect to both axes}. This results in a vanishing total spectral function $\Delta \mathcal{A}_{R,+}=\Delta \mathcal{A}_{R,+}^< +\Delta \mathcal{A}_{R,+}^>=0$.
On the other hand, a {manifest} asymmetry appears in the presence of interactions ($K<1$), where the excess lesser spectral functions for positive and negative charges become independent.
Another clear feature in \Fref{fig:plots-lesser-Rplus}(b) is that $\Delta\mathcal{A}_{R,+}^<(\omega)\propto\Theta(-\omega)$ when $Q_{R,+}=-1$, independently of interactions.
This behavior is uniquely due to the specific shape of integer Lorentzian pulses and, in particular, to the following peculiar property of their photoassisted coefficients:
\begin{equation}
p_\ell(q)=0\qquad\forall\,\ell\,\mathrm{sign}(q)<-|q|\,,\quad q\in\mathbb{Z}\,.
\label{eq:prop-pl}
\end{equation}
\begin{figure}[htbp]
	\begin{center}
		\includegraphics[width=\textwidth]{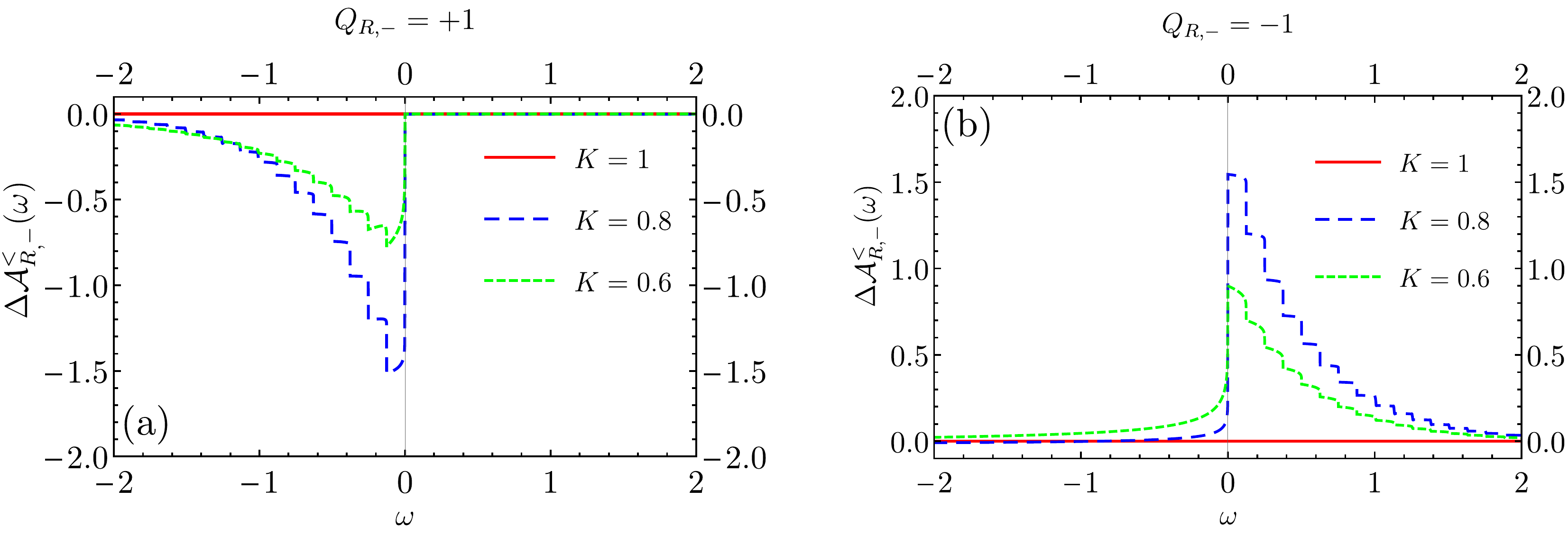}
	\end{center}
	\caption{Lesser spectral function $\Delta\mathcal{A}_{R,-}^<(\omega)$, in units of $w(uT)^{-1}$, as a function of $\omega$, in units of $w^{-1}$. Charges are $Q_{R,-}=\pm 1$, as specified on top of each panel and different values of the interaction strength $K$ are indicated in the label. In all plots we set a representative value for the period $T=50 w$ and $a=0.01uw$.}
	\label{fig:plots-lesser-Rminus}
\end{figure}
In the non-interacting case, $K=1$, the additional symmetry relation \eref{eq:symmetry-k1} is responsible for the appearance of a $\Theta(+\omega)$ also at the positive charge value $Q_{R,+}=+1$. This feature is spoiled when $K<1$, where $\Delta\mathcal{A}_{R,+}^<(\omega)$ is finite for both $\omega\gtrless 0$ [\Fref{fig:plots-lesser-Rplus}(a)].

Importantly, in the non-interacting case only the channels $(R,+)$ and $(L,-)$ have a finite spectral weight, while for $K<1$ other two channels are also present. The presence of these additional contributions in the non-equilibrium spectral function and in its variation are thus a unique fingerprint of interactions. The variation $\Delta \mathcal{A}_{R,-}^<(\omega)$ is shown for different interaction strengths $K<1$ in \Fref{fig:plots-lesser-Rminus}, where, again, the two panels refer to opposite injected integer charges $Q_{R,-}=\pm1$. 
The plots show that $\Delta \mathcal{A}_{R,-}^<(\omega)$ is nonvanishing for both $\omega\gtrless 0$ when its charge is negative, while a $\Theta(-\omega)$ appears for a positive charge. This shows once more that the presence of interactions results in an asymmetric behavior of the non-equilibrium spectral function in response to positive or negative excitations.

{Further information can be obtained from Figures \ref{fig:plots-lesser-Rplus}-\ref{fig:plots-lesser-Rminus} by looking at the sign of the spectral functions.}
We have already pointed out in \eref{eq:sum-rule} that the integral of $\Delta\mathcal{A}_{r,\eta}^<(\omega)$ yields the charge $Q_{r,\eta}$. By inspecting the plots, we see that for the channel $(R,+)$ in \Fref{fig:plots-lesser-Rplus} the sign of the $\mathcal{A}_{R,+}^<$ is everywhere the same as the one of its integral. It is actually possible to prove that the spectral function is always positive-definite for integer $Q_{R,+}$ (see Appendix for details).
This shows that for the channel $(R,+)$ Lorentzian pulses with associated integer charges remain minimal even in the presence of interactions.
On the contrary, for the channel $(R,-)$ the sign of the spectral function is not definite and at low $\omega$ it is actually the opposite of the one required by the sum rule (see Appendix).
For this reason, the function $\Delta\mathcal{A}_{R,-}^<$ is not minimal also in the case of associated integer charges. The main properties of the spectral functions $\Delta\mathcal{A}^<_{R\pm}$ are summarized in \Tref{tab:summary}.
\begin{table}
	\centering
	\begin{tabular}{c|cc}
		\multicolumn{3}{c}{Properties of $\Delta\mathcal{A}^<_{R,+}(\omega)$}\\[.5em]
		\toprule[.2em]
		$Q_{R,+}$&$K=1$&$K<1$\\
		\midrule
		\multirow{2}{*}{$+1$}&$\quad\propto \Theta(+\omega)\quad$&always finite\\
		&minimal&minimal\\[1em]
		\multirow{2}{*}{$-1$}&$\quad\propto \Theta(-\omega)\quad$&$\propto \Theta(-\omega)$\\
		&minimal& minimal\\
		\bottomrule
	\end{tabular}\hspace{1cm}
	\begin{tabular}{c|cc}
		\multicolumn{3}{c}{Properties of $\Delta\mathcal{A}^<_{R,-}(\omega)$}\\[.5em]
		\toprule[.2em]
		$Q_{R,-}$&$\quad K=1$&$K<1$\\
		\midrule
		\multirow{2}{*}{$+1$}&\multirow{2}{*}{$\quad$zero}&$\quad\propto \Theta(-\omega)\quad$\\
		& &non-minimal\\[1em]
		\multirow{2}{*}{$-1$}&\multirow{2}{*}{$\quad$zero}&always finite\\
		& &non-minimal\\
		\bottomrule
	\end{tabular}
	\caption{ Summary of the main properties of the spectral functions for integer charges. In the non-interacting case $K=1$, no charge is injected on the $(R,-)$ channel and therefore $\Delta \mathcal{A}^<_{R,-}(\omega)$ vanishes everywhere. The properties of the other spectral functions can be obtained by using the symmetry relations \eref{eq:relations-spectral}. }
	\label{tab:summary}
\end{table}
\begin{figure}[htbp]
	\begin{center}
		\includegraphics[width=\textwidth]{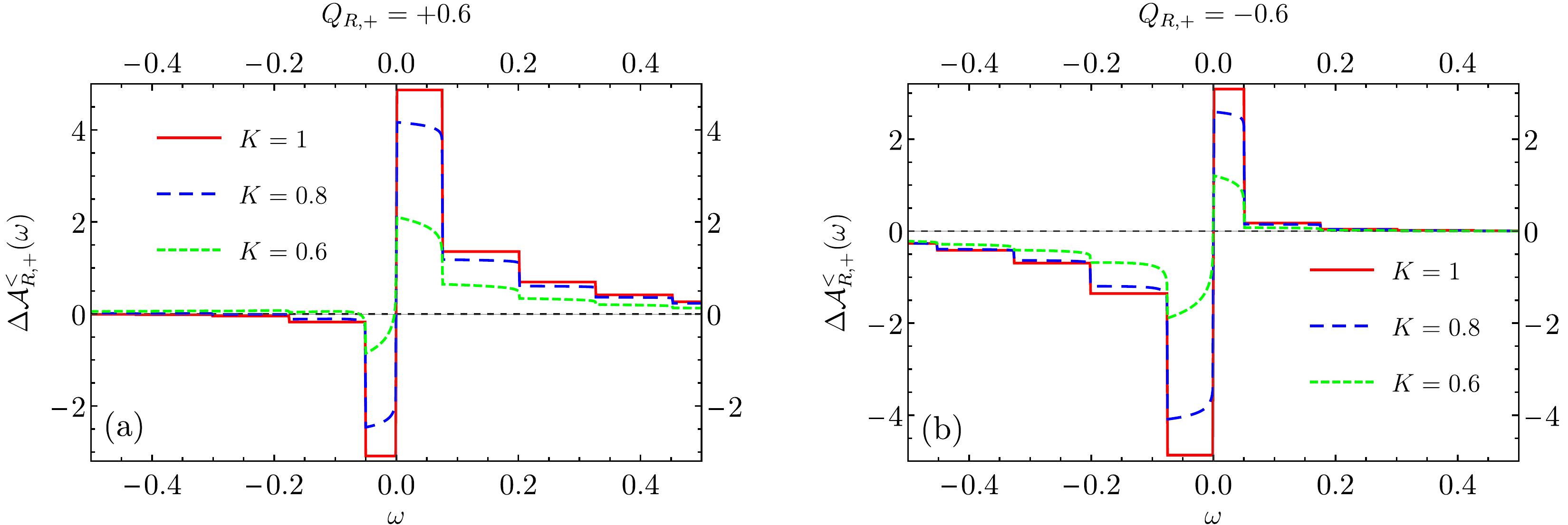}
	\end{center}
	\caption{Lesser spectral fuction $\Delta\mathcal{A}^<_{R,+}(\omega)$, in units of $w(uT)^{-1}$, as a function of $\omega$, in units of $w^{-1}$. Here, we show an example of non-integer charge, with $Q_{R,+}=\pm 0.6$. We set $T=50 w$ and $a=0.01uw$.}
	\label{fig:non-int-q}
\end{figure}

Having described the peculiarities of Lorentzian pulses with associated integer charges $Q_{r,\eta}$, a comment on a generic situation of non-integer charge is in order. In this case qualitative differences appear and have to be considered, since it is in general not possible to have all charges $Q_{r,\eta}$ simultaneously integer, unless for very specific values of the interaction strength. As an example, in \Fref{fig:non-int-q} we plot the function $\Delta\mathcal{A}^<_{R,+}(\omega)$ for $Q_{R,+}=\pm 0.6$, directly obtained from \Eref{eq:lesser-spectral}.
The main difference to be appreciated with respect to the integer case in \Fref{fig:plots-lesser-Rplus} is the absence of the $\Theta(-\omega)$ and that the sign of $\Delta\mathcal{A}^<_{R,+}(\omega)$ is not defined, showing a non-minimal character.

\section{Possible experimental signatures}
\label{sec:tip}
In this Section we show how the intrinsic properties of the spectral functions can be probed by relying on a scanning tunneling setup with a spin-polarized tip, kept at a given (but tunable) bias $V_\mathrm{tip}$ with respect to the helical channels. As a result of this coupling, a tunnel current flows between the tip and the system. The spin polarization of the tip allows us to access all possible channels of the helical liquid, by exploiting the spin-momentum locking~\cite{dasprl2011,calzona15physicaE}. Recently, this technique has been successfully used to probe the surface states of three-dimensional topological insulators~\cite{liu15-tip3dti,hus17-tip3dti,voigtlander18-tip}. We will show how information about the spectral functions can be obtained from the current flowing into the tip and its fluctuations.
\begin{figure}[htbp]
	\begin{center}
		\includegraphics[width=0.65\textwidth]{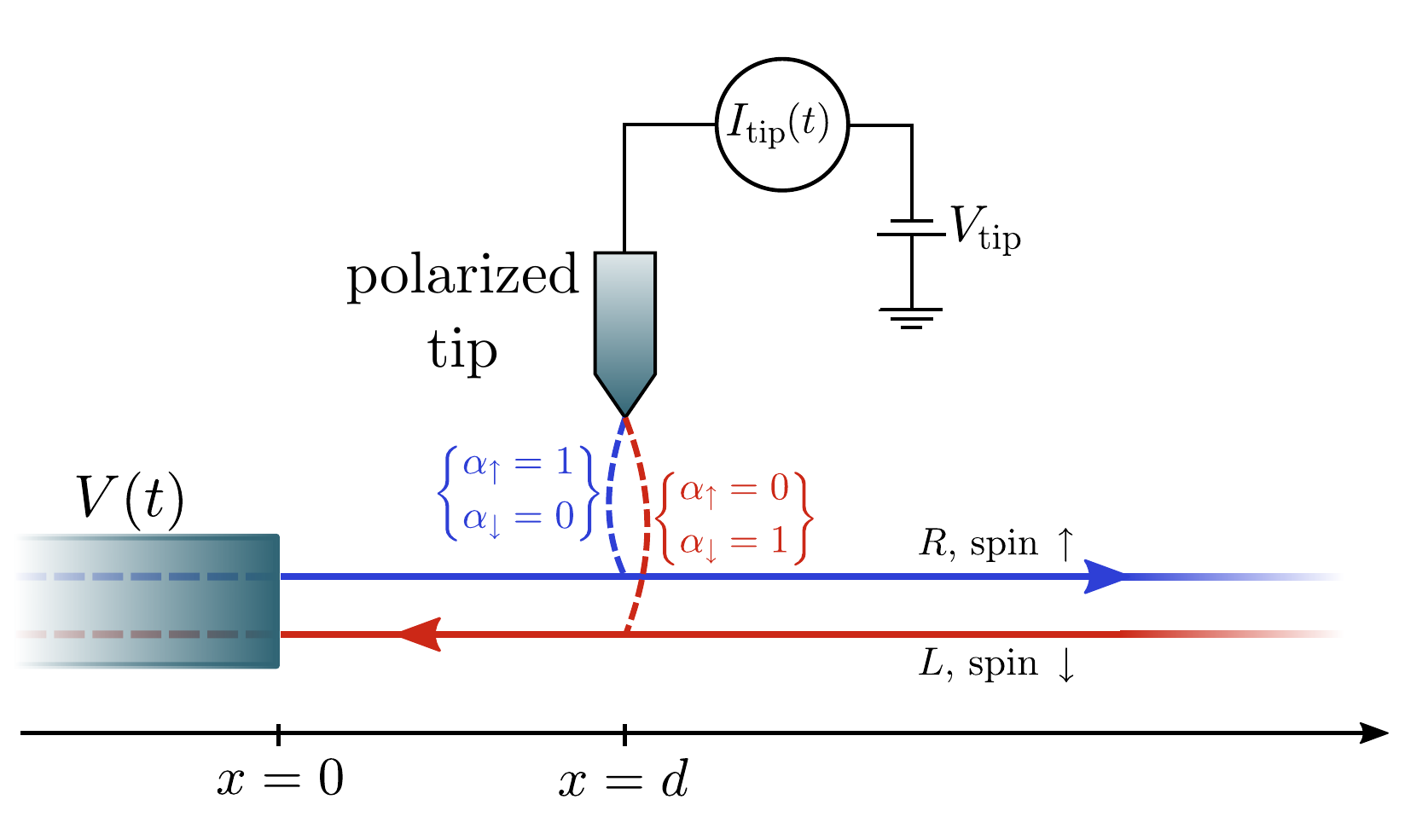}
		\caption{Sketch of the proposed setup. The interacting helical channels are driven by the periodic time-dependent voltage $V(t)$, applied in the region $x<0$. A spin-polarized tip is placed at $d>0$ and allows the spin-preserving tunneling of electrons, selecting their spin according to its polarization. The tip is modeled as a non-interacting system and is biased with a voltage $V_\mathrm{tip}$ with respect to the chemical potential of the helical channels. This setup measures the tunnel current between the system and the tip.}
		\label{fig:tip}
	\end{center}
\end{figure}

We consider a spin-polarized tip, placed at a fixed position $d>0$, as sketched in \Fref{fig:tip}. The tip is coupled to the system via the tunneling Hamiltonian
\begin{equation}
\hat{H}_t=\sum_{\sigma=\uparrow,\downarrow}\left[\lambda\left(\alpha_\uparrow\hat{\Psi}_R^\dagger(d)\hat{\chi}_\uparrow+\alpha_\downarrow\hat{\Psi}_L^\dagger(d)\hat{\chi}_\downarrow\right)+\mathrm{H.c.}\right]\,,
\label{eq:tunnel}
\end{equation}
where $\hat{\chi}_\sigma$ is the annihilation operator of electrons with spin projection $\sigma$ on the tip.
The spin-up polarization of the tip is described by $\alpha_\uparrow=1$ and $\alpha_\downarrow=0$, the spin-down one by $\alpha_\uparrow=0$ and $\alpha_\downarrow=1$.
We only include spin-preserving tunneling, therefore, in the Hamiltonian \eref{eq:tunnel}, $\hat{\chi}_\uparrow$ and $\hat{\chi}_\downarrow$ are only coupled to $\hat{\Psi}_R$ and $\hat{\Psi}_L$, respectively~\cite{dasprl2011}.

The tunnel current flowing between the system and the tip, when the latter is polarized with spin $\uparrow$ or $\downarrow$, can be written as
\begin{equation}
\hat{I}_{\mathrm{tip},R/L}(t)=ie\lambda\hat{\Psi}_{R/L}^\dagger(d,t)\hat{\chi}_{\uparrow/\downarrow}(t)+\mathrm{H.c.}
\end{equation}
The noise associated with its fluctuations is
\begin{equation}
S_{\mathrm{tip},r}=2\int_{-T/2}^{T/2}\frac{\rmd t}{T}\int_{-\infty}^{+\infty}\rmd\tau\Braket{\delta\hat I_{\mathrm{tip},r}\left(t+\frac{\tau}{2}\right)\delta\hat I_{\mathrm{tip},r}\left(t-\frac{\tau}{2}\right)}\,,
\end{equation}
where $\delta\hat{I}_{\mathrm{tip},r}(t)=\hat{I}_{\mathrm{tip},r}(t)-\braket{\hat{I}_{\mathrm{tip},r}(t)}$.

Both quantities $I_{\mathrm{tip},r}=\int_{-T/2}^{T/2}\frac{\rmd t}{T}\braket{\hat{I}_{\mathrm{tip},r}(t)}$ and $S_{\mathrm{tip},r}$ are evaluated to lowest order in the coupling constant $\lambda$, finding
\begin{eqnarray}
\fl I_{\mathrm{tip},r}&=2\pi e|\lambda|^2\int_{-\infty}^{+\infty}\rmd\omega\left[\mathcal{A}_{r}^<(\omega;d)\mathcal{A}_\mathrm{tip}^>(\omega-eV_\mathrm{tip})-\mathcal{A}_{r}^>(\omega;d)\mathcal{A}_\mathrm{tip}^<(\omega-eV_\mathrm{tip})\right],
\label{eq:i-tip1}\\
\fl S_{\mathrm{tip},r}&=4\pi e^2|\lambda|^2\int_{-\infty}^{+\infty}\rmd\omega\left[\mathcal{A}_{r}^<(\omega;d)\mathcal{A}_\mathrm{tip}^>(\omega-eV_\mathrm{tip})+\mathcal{A}_{r}^>(\omega;d)\mathcal{A}_\mathrm{tip}^<(\omega-eV_\mathrm{tip})\right]\,.
\label{eq:s-tip1}
\end{eqnarray}
Here, $\mathcal{A}_r^\gtrless(\omega,d)=\mathcal{A}_0^\gtrless(\omega)+\Delta\mathcal{A}_r^\gtrless(\omega,d)$, with $\Delta\mathcal{A}_r^\gtrless(\omega,d)$ defined in \eref{eq:def-spectral} and the equilibrium term $\mathcal{A}_0^\gtrless(\omega)$ is the Fourier transform of \eref{eq:corr-eq}~\cite{fisher_review}:
\begin{equation}
\mathcal{A}_0^\gtrless(\omega)=\pm\rmi\int_{-\infty}^{+\infty}\frac{\rmd\tau}{2\pi}\rme^{\rmi\omega\tau}G_0^\gtrless(\tau)=\frac{1}{2\pi u}\frac{\rme^{-\frac{a|\omega|}{u}}}{\Gamma(1+2A_-^2)}\Theta(\pm\omega)\left[\frac{a|\omega|}{u}\right]^{2A_-^2}.
\label{eq:spectral-eq}
\end{equation}
Since $d>0$, spectral functions are only related to the excitation on the channel $(r,+)$ [see \Eref{eq:spectral-time-r}]. For notational convenience, we will not include the index $+$ in this Section, since there is no ambiguity.
The equilibrium spectral function of the tip is defined as
\begin{equation}
\mathcal{A}_\mathrm{tip}^\gtrless(\omega)=\pm\frac{\rmi}{2\pi}\int_{-\infty}^{+\infty}\rmd\tau\,G_\mathrm{tip}^\gtrless(\tau)\,\rme^{\rmi\omega\tau}\,,
\end{equation}
where the tip Green functions, $G_\mathrm{tip}^<(\tau)=\rmi\braket{\hat{\chi}_\sigma^\dagger(0)\hat{\chi}_\sigma(\tau)}$ and $G_\mathrm{tip}^>(\tau)=-\rmi\braket{\hat{\chi}_\sigma(0)\hat{\chi}_\sigma^\dagger(-\tau)}$, are independent of the spin $\sigma$.
Since the tip is non-interacting, their expressions are obtained from \eref{eq:spectral-eq}, with $A_-=0$ and $u=\vf$. Then, for $a\to 0$, $\mathcal{A}_\mathrm{tip}^\gtrless(\omega)=(2\pi\vf)^{-1}\Theta(\pm\omega)$. Therefore, the deviations $\Delta I_{\mathrm{tip},r}$ in the current \eref{eq:i-tip1} due to the effect of the drive $V(t)$ only can be expressed as
\begin{equation}
\fl\Delta I_{\mathrm{tip},r}(V_\mathrm{tip})=\frac{e|\lambda|^2}{\vf}\left[\int_{eV_\mathrm{tip}}^{+\infty}\rmd\omega\Delta\mathcal{A}_{r}^<(\omega,q)-\int_{-\infty}^{eV_\mathrm{tip}}\rmd\omega\Delta\mathcal{A}_{r}^<(-\omega,-q)\right].
\label{eq:i-tip2}
\end{equation}
Similarly, the deviations $\Delta S_{\mathrm{tip},r}$ in the noise \eref{eq:s-tip1} read
\begin{equation}
\fl\Delta S_{\mathrm{tip},r}(V_\mathrm{tip})=\frac{2e^2|\lambda|^2}{\vf}\left[\int_{eV_\mathrm{tip}}^{+\infty}\rmd\omega\Delta\mathcal{A}_{r}^<(\omega,q)+\int_{-\infty}^{eV_\mathrm{tip}}\rmd\omega\Delta\mathcal{A}_{r}^<(-\omega,-q)\right].
\label{eq:s-tip2}
\end{equation}
It is worth noticing that $\Delta I_{\mathrm{tip},r}\ne 0$ even at zero static bias $(V_\mathrm{tip}=0)$, because the helical channels are driven out of equilibrium by $V(t)$, which has a non-zero dc component. Moreover, while in general the deviations $\Delta I_{\mathrm{tip},r}$ and $\Delta S_{\mathrm{tip},r}$ are different from $I_{\mathrm{tip},r}$ and $S_{\mathrm{tip},r}$ obtained in \eref{eq:i-tip1} and \eref{eq:s-tip1}, the difference disappears at $V_\mathrm{tip}=0$.
We can now introduce the following excess noises by combining the last two equations:
\begin{equation}
\fl\Delta S_{\mathrm{exc},r}^{(\pm)}(V_\mathrm{tip})=\Delta S_{\mathrm{tip},r}(V_\mathrm{tip})\mp 2e\Delta I_{\mathrm{tip},r}(V_\mathrm{tip})=\pm\frac{4e^2|\lambda|^2}{\vf}\int_{\mp\infty}^{eV_\mathrm{tip}}\rmd\omega\Delta\mathcal{A}_{r}^\gtrless(\omega,q)\,.
\label{eq:delta-s}
\end{equation}
These quantities represent the deviations of the noise from its Poissonian limiting value.
It is now possible to extract information about the intrinsic spectral properties of the helical channels.

\begin{figure}[htbp]
	\begin{center}
		\includegraphics[width=0.5\textwidth]{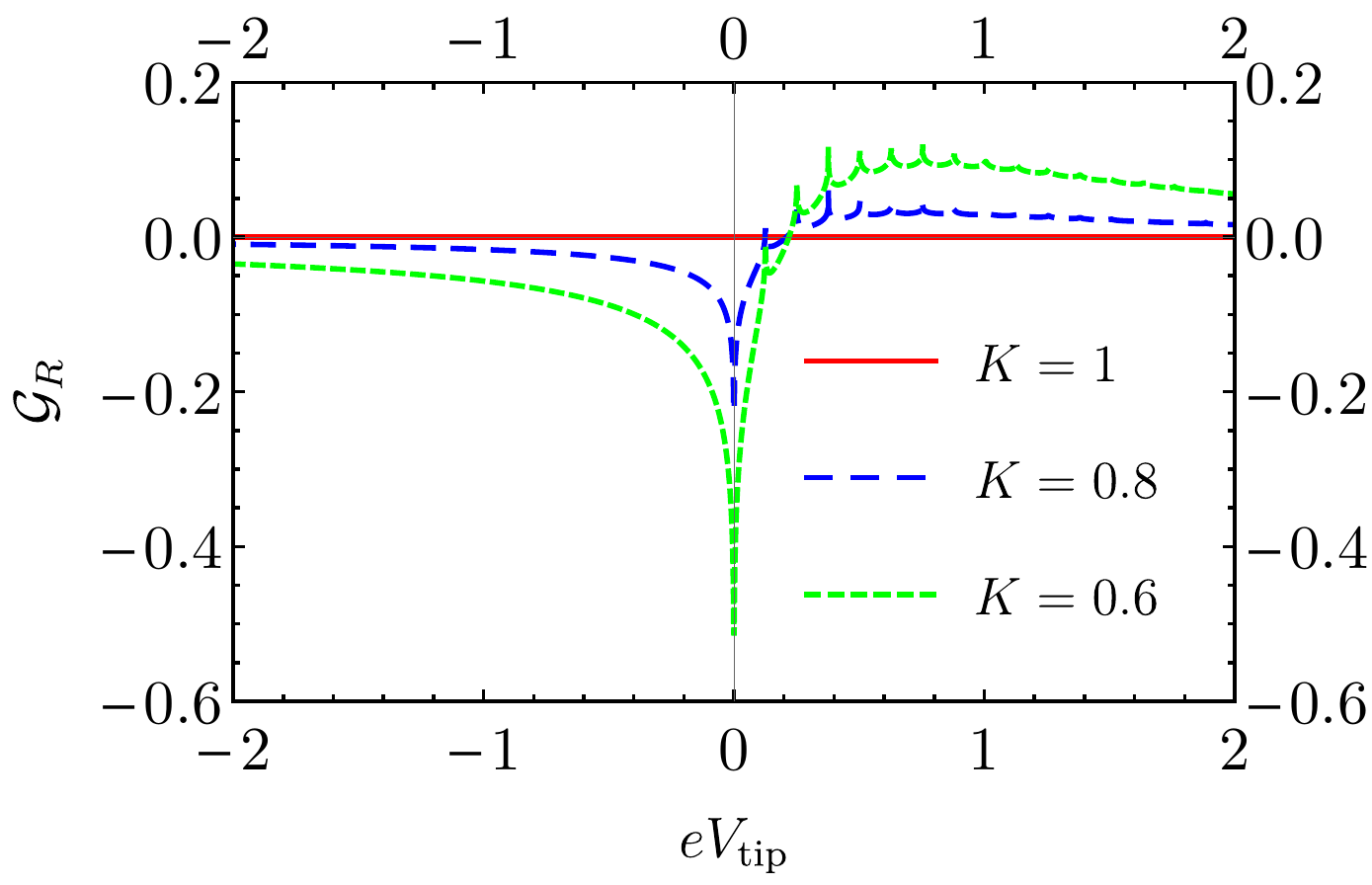}
	\end{center}
	\caption{Excess differential conductance $\mathcal{G}_R=\frac{\partial\Delta I_{\mathrm{tip},R}(V_\mathrm{tip})}{\partial V_\mathrm{tip}}$ in units of $\frac{e^2|\lambda|^2w}{\vf u T}$, as a function of $eV_\mathrm{tip}$, in units of $w^{-1}$. These plots are obtained for a Lorentzian drive with $Q_{R,+}=1$ and $T=50 w$ and directly give, up to a sign, the total excess spectral function $\Delta\mathcal{A}_{R}(\omega)$, as established by \Eref{eq:diff-cond}. Notice that in the non-interacting case, the result is zero, due to the symmetry \eref{eq:symmetry-k1} of the spectral functions.}
	\label{fig:diff-cond}
\end{figure}

First of all, the variation of the total spectral distribution $\Delta\mathcal{A}_{r}(\omega)=\Delta\mathcal{A}_{r}^<(\omega)+\Delta\mathcal{A}_{r}^>(\omega)$ can be obtained from the excess differential conductance, namely
\begin{equation}
\Delta\mathcal{A}_{r}(\omega)=-\frac{\vf}{e^2|\lambda|^2}\left.\frac{\partial\Delta I_{\mathrm{tip},r}(V_\mathrm{tip})}{\partial V_\mathrm{tip}}\right|_{eV_\mathrm{tip}=\omega}\,.
\label{eq:diff-cond}
\end{equation}
Given \Eref{eq:total-spectral}, in a non-interacting system $\Delta\mathcal{A}_{r}(\omega)=0$, regardless the shape of the drive. This does not hold anymore as soon as interactions are present. Indeed, by considering the case of a Lorentzian drive, we show in \Fref{fig:diff-cond} the variation of the excess differential conductance for different values of the interaction strength, in the case where the tip is polarized with $\sigma=\uparrow$. Thanks to the sharply different behavior between interacting and non-interacting case, it is possible from a measurement of the current $\Delta I_{\mathrm{tip},r}$ to probe whether the system is interacting or not.

Additional information can be obtained by taking the derivative of the two excess noises introduced in \eref{eq:delta-s}:
\begin{equation}
\Delta\mathcal{A}_{r}^\gtrless(\omega,q)=\pm\frac{\vf}{4e^3|\lambda|^2}\left.\frac{\partial \Delta S_{\mathrm{exc},r}^{(\pm)}(V_\mathrm{tip})}{\partial V_\mathrm{tip}}\right|_{eV_\mathrm{tip}=\omega}\,.
\label{eq:spectral-reconstruction}
\end{equation}
This relation makes it possible, by varying $V_\mathrm{tip}$, to reconstruct both the greater and lesser spectral functions and access all the features presented in the previous Section. Notice also that spectral functions of both channels $R$ and $L$ can be investigated by simply changing the polarization of the tip.

{Further information about the excess noise can be extracted from \eref{eq:delta-s}}. At $V_\mathrm{tip}=0$, the two quantities $\Delta S_{\mathrm{exc},R}^{(\pm)}(0)$ vanish when the excitation on the channel $(R,+)$ is an integer charge with Lorentzian shape. In particular $\Delta S_{\mathrm{exc},R}^{(+)}(0)=0$ when $Q_{R,+}$ is a positive integer (electron-like excitation, with $q>0$), while $\Delta S_{\mathrm{exc},R}^{(-)}(0)=0$ when $Q_{R,+}$ is a negative integer (hole-like excitation, with $q<0$).
This is due to the fact that $\Delta\mathcal{A}_R^>(\omega)|_{Q_{R,+}=+1}\propto\Theta(\omega)$, while $\Delta\mathcal{A}_R^<(\omega)|_{Q_{R,+}=-1}\propto\Theta(-\omega)$ (see \Fref{fig:plots-lesser-Rplus}). Through the same reasoning we see that $\Delta S_{\mathrm{exc},L}^{(\pm)}(0)=0$ when $Q_{L,+}$ is a negative [$q>0$, see \eref{eq:charge}] or positive $(q<0)$ integer, respectively. Let us focus on $\Delta S_{\mathrm{exc},r}^{(+)}(0)$ and analyze the conditions for it to vanish. When $r=R$, we need $Q_{R,+}=n$, with $n\in\mathbb{N}^+$. In terms of the initial injected value $q$, this means
\begin{equation}
q=\frac{2n}{1+K}\,.
\label{eq:qR}
\end{equation}
Likewise, when $r=L$, we need $Q_{L,+}=-m$, with $m\in\mathbb{N}^+$, namely
\begin{equation}
q=\frac{2m}{1-K}\,.
\label{eq:qL}
\end{equation}
\begin{figure}[htbp]
	\begin{center}
		\includegraphics[width=\textwidth]{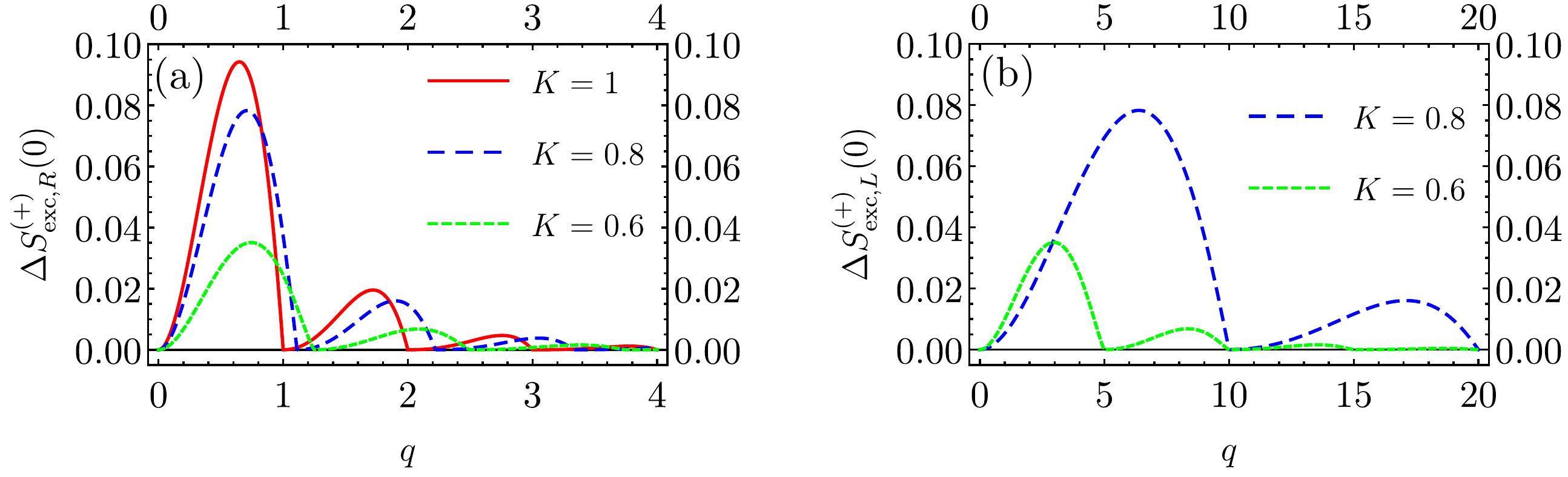}
	\end{center}
	\caption{(a): excess noise $\Delta S_{\mathrm{exc},R}^{(+)}(0)$, in units of $S_0=e^2\Omega|\lambda|^2/(\pi u^2)$, as a function of $q$, for different values of the interaction $K$. The zeros are located at the points given by \eref{eq:qR}, namely when $Q_{R,+}$ is a positive integer. (b): excess noise $\Delta S_{\mathrm{exc},L}^{(+)}(0)$, in units of $S_0$. Here, the zeros are located according to \eref{eq:qL}. Note that no signal is present at $K=1$. In both panels we set $w=0.1T$ and $u(a\Omega)^{-1}=10$. }
	\label{fig:excess-noise}
\end{figure}
In \Fref{fig:excess-noise} we present the behavior of excess noise $\Delta S_{\mathrm{exc},R/L}^{(+)}(0)$ as a function of the experimentally tunable parameter $q$, showing that the zeros are indeed located at the points given by \eref{eq:qR} and \eref{eq:qL}. By varying the interaction, the zeros in panel (a) move to higher values of $q$, as required by \Eref{eq:qR}, while the opposite is true in panel (b), according to \Eref{eq:qL}. In the latter case we do not have any signal at $K=1$, since the spectral function on the $L$ channel vanishes. This discussion demonstrates that a measurement of the excess noise could be used to extract the value of the interaction strength. Indeed, by looking for instance at the value of $q$ at which the $n$-th zero in \Fref{fig:excess-noise}(a) occurs, the Luttinger parameter can be determined by solving \Eref{eq:qR} for $K$.

As it is clear from the above discussion, a vanishing excess noise is only due to the presence of proper $\Theta$ functions in the spectral functions. This property is uniquely determined by the Lorentzian drive and is therefore robust with respect to the presence of interactions. Indeed a vanishing excess noise can be achieved at any interaction strength, provided that the conditions in \eref{eq:qR} or \eref{eq:qL} are met.
We recall that a vanishing excess noise in the case of Lorentzian pulses producing excitations with integer charge has been already reported in a QPC geometry for non-interacting systems~\cite{keeling06,dubois13prb,dubois2013levitonsNature} as well as in the integer~\cite{grenier13,acciai18} and fractional quantum Hall effect~\cite{rech16prl}.
Indeed, one can recognize that expressions for the excess noise of a QPC are equivalent to the ones in \eref{eq:i-tip1} and \eref{eq:s-tip1}.
We emphasize however that interactions in counterpropagating helical channels result in a richer phenomenology in the excess noise, as the positions of the zeros depend on $K$.

As discussed in \Sref{sec:state-of-the-art} and \Sref{sec:spectral}, in a non-interacting system a vanishing excess noise directly implies that $\Delta\mathcal{A}_R^<(\omega)$ has a definite sign. This is not anymore true if $K\ne 1$. As an example, when a Lorentzian pulse with $Q_{L,+}$ a negative integer is generated on the channel $(L,+)$, both $\Delta\mathcal{A}_{L,+}^\gtrless(\omega)$ do not have a definite sign, but still $\Delta S_{\mathrm{exc},L}^{(+)}(0)=0$, as we see in \Fref{fig:excess-noise}. We conclude that, apart from the case $K=1$, a vanishing excess noise is not necessarily related to a minimal spectral function (in the sense of absence of additional positive/negative charge). This fact was already noticed in a different context~\cite{vanevic07,belzig16physe}.
One of the striking results of the latter analysis is that contrary to common belief that minimal wave packets need to bear an integer charge, in strongly
correlated systems such as a quantum spin Hall Luttinger liquid system, the excess noise vanishes for non integer charges. 

\section{Conclusions}
\label{sec:concl}
We have analized the non-equilibrium spectral properties of interacting helical channels
in the presence of a time-dependent drive.
In order to better elucidate the effects induced by e-e interactions, we have focused on
the case of a periodic train of integer Lorentzian pulses, which is known to generate minimal excitations in a 1D free-fermion system. We have shown that peculiar asymmetries, related to the sign of the injected charge, appear as a function of the interaction strength {(see Figures \ref{fig:plots-lesser-Rplus} and \ref{fig:plots-lesser-Rminus} and Table \ref{tab:summary})}. Moreover,
the concept of minimal excitations has to be properly considered and is no more directly
related to the vanishing of the excess noise as in the case of free fermions.
These findings can be tested by looking at the tunneling current and its 
fluctuations through a polarized tip, which allows for a spectroscopic investigation of the intrinsic spectral properties of all counterpropagating channels {[see \Eref{eq:spectral-reconstruction}]}.
The spin-polarized tip experiment suggested in this work shows that in correlated electron systems such as the quantum
spin Hall Luttinger liquid, the transmitted charges per period associated with the voltage pulse minimizing the excess noise deviate from integer values {[Equations \eref{eq:qR} and \eref{eq:qL}]}.
This is a direct consequence of the fact that the Luttinger parameter $K$ is smaller than one. Therefore, our analysis can be considered as a novel diagnosis
for detecting fractional charges in quantum spin Hall Luttinger liquids in the presence of time-dependent drives.

\ack
This work was granted access to the HPC resources of Aix-Marseille Universit\'e financed by the project Equip@Meso (Grant No. ANR-10-EQPX29-01). It has been carried out in the framework of project "one shot
reloaded" (Grant No. ANR-14-CE32-0017) and benefited from the support of the Labex ARCHIMEDE (Grant No.
ANR11-LABX-0033) and the AMIDEX project (Grant No. ANR-11-IDEX-0001-02), funded by the "investissements
d'avenir" French Government program managed by the French National Research Agency (ANR).
A. C. acknowledges support by the W\"urzburg-Dresden Cluster of Excellence on Complexity and Topology in Quantum Matter (EXC 2147, project-id 39085490).
M. C. acknowledges support from the Quant-EraNet project Supertop.

\appendix
\section{Sign of the spectral functions}
\label{sec:app}

Here, we show that the function $\Delta\mathcal{A}^<_{R,+}(\omega)$ has a definite sign when evaluated for integer Levitons. For simplicity, it is convenient to consider the case of very long period $T\ggg w$ and thus focus on a single Lorentzian pulse, given in \eref{eq:single-pulse}. In this case, the phase factor \eref{eq:phase} becomes
\begin{equation}
\Xi_{r,\eta}(t,\tau)=\left[\frac{w+\rmi(t-\tau/2)}{w-\rmi(t-\tau/2)}\right]^{\eta\vartheta_r{Q}_{r,\eta}}\left[\frac{w+\rmi(t+\tau/2)}{w-\rmi(t+\tau/2)}\right]^{-\eta\vartheta_r{Q}_{r,\eta}}-1\,.
\end{equation}
If $Q_{r,\eta}$ is integer, further analytical evaluation is possible. Let us focus on the case where $Q_{r,\eta}=n>0$. It is then possible to show that~\cite{moskalets15,glattli16wavepackets,ronetti18crystallization,acciai18}
\begin{equation}
\fl\left[\frac{w+\rmi(t-\tau/2)}{w-\rmi(t-\tau/2)}\right]^n\left[\frac{w-\rmi(t+\tau/2)}{w+\rmi(t+\tau/2)}\right]^n-1=-2\pi\rmi\tau\sum_{j=1}^{n}\chi_j\left(t+\frac{\tau}{2}\right)\chi_j^*\left(t-\frac{\tau}{2}\right)\,,
\end{equation}
with
\begin{equation}
\chi_j(t)=\sqrt{\frac{w}{\pi}}\frac{(t+iw)^{j-1}}{(t-iw)^j}\,.
\end{equation}
By relying on these results, the variation of the spectral function becomes
\begin{equation}
\fl\Delta\tilde\mathcal{A}^<_{R,+}(\omega)=\frac{1}{2\pi u}\int_{-\infty}^{+\infty}\rmd\tau\,\rme^{\rmi\omega\tau}\left[\frac{a}{a-\rmi u\tau}\right]^{2A_-^2}\int_{-\infty}^{+\infty}\rmd t\sum_{j=1}^{n}\chi_j\left(t+\frac{\tau}{2}\right)\chi_j^*\left(t-\frac{\tau}{2}\right)\,.
\end{equation}
Here, the tilde indicates that we adapted the definition of the spectral function to a single-pulse drive, by replacing the integral $T^{-1}\int_{-T/2}^{T/2}\rmd t$ with $\int_{-\infty}^{+\infty}\rmd t$. By evaluating the previous integral we arrive at the following result:
\begin{equation}
\fl\Delta\tilde\mathcal{A}^<_{R,+}(\omega)=\frac{2w}{u}\left(\frac{a}{u}\right)^{2A_-^2}\frac{\rme^{a\omega/u}}{\Gamma(2A_-^2)}\sum_{j=1}^{n}\left[\Theta(\omega)\mathcal{I}_j(\omega,\omega)+\Theta(-\omega)\mathcal{I}_j(\omega,0)\right]\,,
\end{equation}
where $\Gamma$ is the Gamma function, $L_j$ the Laguerre polynomial of order $j$ and
\begin{equation}
\mathcal{I}_j(\omega_1,\omega_2)=\int_{\omega_2}^{+\infty}\rmd\epsilon\,\rme^{-2\epsilon w}\rme^{-\epsilon a/u}|\omega_1-\epsilon|^{2A_-^2-1}L_{j-1}^2(2\epsilon w)\,.
\end{equation}
This results shows that the spectral function $\Delta\mathcal{A}_{R,+}^<$ is always positive and nonvanishing for both $\omega\gtrless 0$, precisely as observed in the main text.

When $Q_{R,+}=m<0$, by following the same steps as before, we find:
\begin{equation}
\Delta\mathcal{A}^<_{R,+}(\omega)=-\frac{2w}{u}\left(\frac{a}{u}\right)^{2A_-^2}\frac{\rme^{a\omega/u}}{\Gamma(2A_-^2)}\sum_{j=1}^{|m|}\Theta(-\omega)\mathcal{J}_j(\omega)\,,
\end{equation}
with
\begin{equation}
\mathcal{J}_j(\omega)=\int_0^{-\omega}\rmd\epsilon\,\rme^{-2\epsilon w}\rme^{\epsilon a/u}L_{j-1}^{2}(2\epsilon w)|\omega+\epsilon|^{2A_-^2-1}\,.
\end{equation}
This shows that, in this case, the spectral function is always negative and vanishes for $\omega>0$.

\begin{figure}[hbtp]
	\begin{center}
		\includegraphics[width=\textwidth]{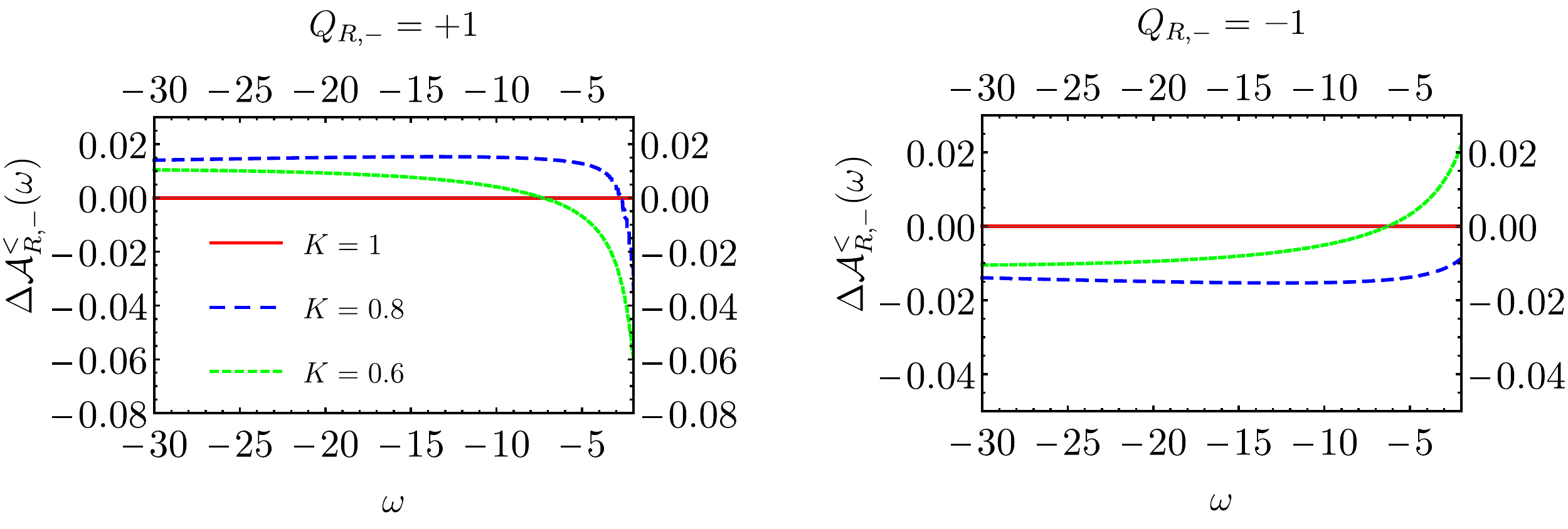}
	\end{center}
	\caption{Behavior of the tails of the spectral function $\Delta\mathcal{A}_{R,-}^<(\omega)$, for $Q_{R,-}=\pm1$. Different values of $K$ are indicated in the left panel. We clearly see a change of sign with respect to what happens at values of $\omega$ closer to zero, see \Fref{fig:plots-lesser-Rminus}. The units on the axes are the same as in \Fref{fig:plots-lesser-Rminus}.}
	\label{fig:tails}
\end{figure}

Concerning the spectral function $\Delta\mathcal{A}_{R,-}^<(\omega)$, we show in \Fref{fig:tails} that its sign is not definite and that a change of sign occurs in the high-energy tail at $\omega<0$, ensuring that the sum rule \eref{eq:sum-rule} is satisfied.
\section*{References}

\begin{thebibliography}{100}
	\expandafter\ifx\csname url\endcsname\relax
	\def\url#1{{\tt #1}}\fi
	\expandafter\ifx\csname urlprefix\endcsname\relax\def\urlprefix{URL }\fi
	\providecommand{\eprint}[2][]{\url{#2}}
	
	\bibitem{nayak08}
	Nayak C, Simon S~H, Stern A, Freedman M and Das~Sarma S 2008 {\em Rev. Mod.
		Phys.\/} {\bf 80} 1083
	
	\bibitem{manchon15}
	Manchon A, Koo H~C, Nitta J, Frolov S~M and Duine R~A 2015 {\em Nature
		Materials\/} {\bf 14} 871
	
	\bibitem{wendin2017}
	Wendin G 2017 {\em Reports on Progress in Physics\/} {\bf 80} 106001
	
	\bibitem{haldane2017nobel}
	Haldane F~D~M 2017 {\em Rev. Mod. Phys.\/} {\bf 89} 040502
	
	\bibitem{klitzing80}
	Klitzing K~v, Dorda G and Pepper M 1980 {\em Phys. Rev. Lett.\/} {\bf 45} 494
	
	\bibitem{girvin1990}
	Prange R~E and Girvin S~M 1990 {\em The Quantum Hall Effect\/}
	(Springer-Verlag)
	
	\bibitem{stern-review}
	Stern A 2008 {\em Annals of Physics\/} {\bf 323} 204
	
	\bibitem{haldane1988prl}
	Haldane F~D~M 1988 {\em Phys. Rev. Lett.\/} {\bf 61} 2015
	
	\bibitem{bhz06}
	Bernevig B~A, Hughes T~L and Zhang S~C 2006 {\em Science\/} {\bf 314} 1757
	
	\bibitem{wu06}
	Wu C, Bernevig B~A and Zhang S~C 2006 {\em Phys. Rev. Lett.\/} {\bf 96} 106401
	
	\bibitem{konig07qsh}
	K{\"o}nig M, Wiedmann S, Br{\"u}ne C, Roth A, Buhmann H, Molenkamp L~W, Qi X~L
	and Zhang S~C 2007 {\em Science\/} {\bf 318} 766
	
	\bibitem{hasan2010colloquium}
	Hasan M~Z and Kane C~L 2010 {\em Rev. Mod. Phys.\/} {\bf 82} 3045
	
	\bibitem{qi2011topological}
	Qi X~L and Zhang S~C 2011 {\em Rev. Mod. Phys.\/} {\bf 83} 1057
	
	\bibitem{dolcetto16review}
	{Dolcetto} G, {Sassetti} M and {Schmidt} T~L 2016 {\em Riv. Nuovo Cimento\/}
	{\bf 39} 113
	
	\bibitem{grenier11}
	Grenier C, Herv\'e R, F\`eve G and Degiovanni P 2011 {\em Mod. Phys. Lett. B\/}
	{\bf 25} 1053
	
	\bibitem{bocquillon12eqo}
	Bocquillon E, Parmentier F~D, Grenier C, Berroir J~M, Degiovanni P, Glattli
	D~C, Pla\ifmmode~\mbox{\c{c}}\else \c{c}\fi{}ais B, Cavanna A, Jin Y and
	F\`eve G 2012 {\em Phys. Rev. Lett.\/} {\bf 108} 196803
	
	\bibitem{bocquillon14}
	Bocquillon E, Freulon V, Parmentier F~D, Berroir J, Pla\c{c}ais B, Wahl C, Rech
	J, Jonckheere T, Martin T, Grenier C, Ferraro D, Degiovanni P and F\`eve G
	2014 {\em Ann. Phys.\/} {\bf 526} 1
	
	\bibitem{bauerle2018review}
	B\"auerle C, Glattli D~C, Meunier T, Portier F, Roche P, Roulleau P, Takada S
	and Waintal X 2018 {\em Rep. Progr. Phys.\/} {\bf 81} 056503
	
	\bibitem{feve07}
	F{\`e}ve G, Mah{\'e} A, Berroir J~M, Kontos T, Pla{\c c}ais B, Glattli D~C,
	Cavanna A, Etienne B and Jin Y 2007 {\em Science\/} {\bf 316} 1169
	
	\bibitem{mahe2010}
	Mah\'e A, Parmentier F~D, Bocquillon E, Berroir J~M, Glattli D~C, Kontos T,
	Pla\ifmmode~\mbox{\c{c}}\else \c{c}\fi{}ais B, F\`eve G, Cavanna A and Jin Y
	2010 {\em Phys. Rev. B\/} {\bf 82} 201309
	
	\bibitem{dubois2013levitonsNature}
	Dubois J, Jullien T, Portier F, Roche P, Cavanna A, Jin Y, Wegscheider W,
	Roulleau P and Glattli D~C 2013 {\em Nature\/} {\bf 502} 659
	
	\bibitem{bocquillon13homscience}
	Bocquillon E, Freulon V, Berroir J~M, Degiovanni P, Pla{\c c}ais B, Cavanna A,
	Jin Y and F{\`e}ve G 2013 {\em Science\/} {\bf 339} 1054
	
	\bibitem{bocquillon2013}
	Bocquillon E, Freulon V, Berroir J~M, Degiovanni P, Pla\c{c}ais B, Cavanna A,
	Jin Y and F\`eve G 2013 {\em Nature Communications\/} {\bf 4} 1839
	
	\bibitem{freulon15hom}
	Freulon V, Marguerite A, Berroir J~M, Pla{\c c}ais B, Cavanna A, Jin Y and
	F\`eve G 2015 {\em Nature Communications\/} {\bf 6} 6854
	
	\bibitem{neder06}
	Neder I, Heiblum M, Levinson Y, Mahalu D and Umansky V 2006 {\em Phys. Rev.
		Lett.\/} {\bf 96} 016804
	
	\bibitem{roulleau07MZ}
	Roulleau P, Portier F, Glattli D~C, Roche P, Cavanna A, Faini G, Gennser U and
	Mailly D 2007 {\em Phys. Rev. B\/} {\bf 76} 161309
	
	\bibitem{chalker2007MZ}
	Chalker J~T, Gefen Y and Veillette M~Y 2007 {\em Phys. Rev. B\/} {\bf 76}
	085320
	
	\bibitem{levkivskyi08modelnu2}
	Levkivskyi I~P and Sukhorukov E~V 2008 {\em Phys. Rev. B\/} {\bf 78} 045322
	
	\bibitem{kovrizhin2009MZ}
	Kovrizhin D~L and Chalker J~T 2009 {\em Phys. Rev. B\/} {\bf 80} 161306
	
	\bibitem{parmentier12mesoscopic}
	Parmentier F~D, Bocquillon E, Berroir J~M, Glattli D~C,
	Pla\ifmmode~\mbox{\c{c}}\else \c{c}\fi{}ais B, F\`eve G, Albert M, Flindt C
	and B\"uttiker M 2012 {\em Phys. Rev. B\/} {\bf 85} 165438
	
	\bibitem{wahl14prl}
	Wahl C, Rech J, Jonckheere T and Martin T 2014 {\em Phys. Rev. Lett.\/} {\bf
		112} 046802
	
	\bibitem{ferraro2014noise}
	Ferraro D, Carrega M, Braggio A and Sassetti M 2014 {\em New Journal of
		Physics\/} {\bf 16} 043018
	
	\bibitem{ferraro15antidot}
	Ferraro D, Rech J, Jonckheere T and Martin T 2015 {\em Phys. Rev. B\/} {\bf 91}
	205409
	
	\bibitem{tewari2016}
	Tewari S, Roulleau P, Grenier C, Portier F, Cavanna A, Gennser U, Mailly D and
	Roche P 2016 {\em Phys. Rev. B\/} {\bf 93} 035420
	
	\bibitem{guiducci19}
	Guiducci S, Carrega M, Biasiol G, Sorba L, Beltram F and Heun S 2019 {\em
		physica status solidi (RRL)\/} {\bf 13} 1800222
	
	\bibitem{glattli16wavepackets}
	Glattli D and Roulleau P 2016 {\em Physica E: Low-dimensional Systems and
		Nanostructures\/} {\bf 76} 216
	
	\bibitem{dolcini11}
	Dolcini F 2011 {\em Phys. Rev. B\/} {\bf 83} 165304
	
	\bibitem{hofer2013}
	Hofer P~P and B\"uttiker M 2013 {\em Phys. Rev. B\/} {\bf 88} 241308
	
	\bibitem{inhofer2013}
	Inhofer A and Bercioux D 2013 {\em Phys. Rev. B\/} {\bf 88} 235412
	
	\bibitem{ferraro14HOMtopo}
	Ferraro D, Wahl C, Rech J, Jonckheere T and Martin T 2014 {\em Phys. Rev. B\/}
	{\bf 89} 075407
	
	\bibitem{calzona15physicaE}
	{Calzona} A, {Carrega} M, {Dolcetto} G and {Sassetti} M 2015 {\em Physica E
		Low-Dimensional Systems and Nanostructures\/} {\bf 74} 630
	
	\bibitem{dolcetto16entanglement}
	Dolcetto G and Schmidt T~L 2016 {\em Phys. Rev. B\/} {\bf 94} 075444
	
	\bibitem{calzona16energypart}
	Calzona A, Acciai M, Carrega M, Cavaliere F and Sassetti M 2016 {\em Phys. Rev.
		B\/} {\bf 94} 035404
	
	\bibitem{ronetti16}
	Ronetti F, Vannucci L, Dolcetto G, Carrega M and Sassetti M 2016 {\em Phys.
		Rev. B\/} {\bf 93} 165414
	
	\bibitem{ronetti17polarized}
	Ronetti F, Carrega M, Ferraro D, Rech J, Jonckheere T, Martin T and Sassetti M
	2017 {\em Phys. Rev. B\/} {\bf 95} 115412
	
	\bibitem{acciai17}
	Acciai M, Calzona A, Dolcetto G, Schmidt T~L and Sassetti M 2017 {\em Phys.
		Rev. B\/} {\bf 96} 075144
	
	\bibitem{bendias18}
	Bendias K, Shamim S, Herrmann O, Budewitz A, Shekhar P, Leubner P, Kleinlein J,
	Bocquillon E, Buhmann H and Molenkamp L~W 2018 {\em Nano Letters\/} {\bf 18}
	4831
	
	\bibitem{strunz19}
	{Strunz} J, {Wiedenmann} J, {Fleckenstein} C, {Lunczer} L, {Beugeling} W,
	{M{\"u}ller} V~L, {Shekhar} P, {Traverso Ziani} N, {Shamim} S and {Kleinlein}
	J 2019  arXiv:1905.08175
	
	\bibitem{dubois13prb}
	Dubois J, Jullien T, Grenier C, Degiovanni P, Roulleau P and Glattli D~C 2013
	{\em Phys. Rev. B\/} {\bf 88} 085301
	
	\bibitem{grenier13}
	Grenier C, Dubois J, Jullien T, Roulleau P, Glattli D~C and Degiovanni P 2013
	{\em Phys. Rev. B\/} {\bf 88} 085302
	
	\bibitem{ferraro13}
	Ferraro D, Feller A, Ghibaudo A, Thibierge E, Bocquillon E, F\`eve G, Grenier C
	and Degiovanni P 2013 {\em Phys. Rev. B\/} {\bf 88} 205303
	
	\bibitem{jullien14tomography}
	Jullien T, Roulleau P, Roche B, Cavanna A, Jin Y and Glattli D~C 2014 {\em
		Nature\/} {\bf 514} 603
	
	\bibitem{moskalets15}
	Moskalets M 2015 {\em Phys. Rev. B\/} {\bf 91} 195431
	
	\bibitem{moskalets16}
	Moskalets M 2016 {\em Phys. Rev. Lett.\/} {\bf 117} 046801
	
	\bibitem{rech16prl}
	Rech J, Ferraro D, Jonckheere T, Vannucci L, Sassetti M and Martin T 2017 {\em
		Phys. Rev. Lett.\/} {\bf 118} 076801
	
	\bibitem{dolcini16chiral}
	Dolcini F, Iotti R~C, Montorsi A and Rossi F 2016 {\em Phys. Rev. B\/} {\bf 94}
	165412
	
	\bibitem{dolcini17}
	Dolcini F 2017 {\em Phys. Rev. B\/} {\bf 95} 085434
	
	\bibitem{vannucci17heat}
	Vannucci L, Ronetti F, Rech J, Ferraro D, Jonckheere T, Martin T and Sassetti M
	2017 {\em Phys. Rev. B\/} {\bf 95} 245415
	
	\bibitem{ronetti18crystallization}
	Ronetti F, Vannucci L, Ferraro D, Jonckheere T, Rech J, Martin T and Sassetti M
	2018 {\em Phys. Rev. B\/} {\bf 98} 075401
	
	\bibitem{acciai18}
	Acciai M, Carrega M, Rech J, Jonckheere T, Martin T and Sassetti M 2018 {\em
		Phys. Rev. B\/} {\bf 98} 035426
	
	\bibitem{ferraro2018review}
	Ferraro D, Ronetti F, Vannucci L, Acciai M, Rech J, Jonckheere T, Martin T and
	Sassetti M 2018 {\em The European Physical Journal Special Topics\/} {\bf
		227} 1345
	
	\bibitem{dolcini18}
	Dolcini F and Rossi F 2018 {\em The European Physical Journal Special Topics\/}
	{\bf 227} 1323
	
	\bibitem{levitov96}
	Levitov L~S, Lee H and Lesovik G~B 1996 {\em J. Math. Phys.\/} {\bf 37}
	4845--4866
	
	\bibitem{levitov97}
	Ivanov D~A, Lee H~W and Levitov L~S 1997 {\em Phys. Rev. B\/} {\bf 56}
	6839--6850
	
	\bibitem{keeling06}
	Keeling J, Klich I and Levitov L~S 2006 {\em Phys. Rev. Lett.\/} {\bf 97}
	116403
	
	\bibitem{battista14}
	Battista F, Haupt F and Splettstoesser J 2014 {\em Phys. Rev. B\/} {\bf 90}
	085418
	
	\bibitem{forrester14}
	Forrester D~M and Kusmartsev F~V 2014 {\em Nanoscale\/} {\bf 6} 7594
	
	\bibitem{forrester15}
	Forrester D~M 2015 {\em RSC Adv.\/} {\bf 5} 5442
	
	\bibitem{dasenbrook15}
	Dasenbrook D and Flindt C 2015 {\em Phys. Rev. B\/} {\bf 92}(16) 161412
	
	\bibitem{dasenbrook16}
	Dasenbrook D and Flindt C 2016 {\em Phys. Rev. B\/} {\bf 93} 245409
	
	\bibitem{moskalets16physe}
	Moskalets M and Haack G 2016 {\em Physica E: Low-dimensional Systems and
		Nanostructures\/} {\bf 82} 204
	
	\bibitem{suzuki17}
	Suzuki T~J 2017 {\em Phys. Rev. B\/} {\bf 95} 241302
	
	\bibitem{hofer17}
	Hofer P~P, Dasenbrook D and Flindt C 2017 {\em physica status solidi (b)\/}
	{\bf 254} 1600582
	
	\bibitem{moskalets17-pss}
	Moskalets M and Haack G 2017 {\em physica status solidi (b)\/} {\bf 254}
	1600616
	
	\bibitem{cabart18}
	Cabart C, Roussel B, F\`eve G and Degiovanni P 2018 {\em Phys. Rev. B\/} {\bf
		98} 155302
	
	\bibitem{moskalets18}
	Moskalets M 2018 {\em Phys. Rev. B\/} {\bf 97} 155411
	
	\bibitem{dashti19}
	Dashti N, Misiorny M, Kheradsoud S, Samuelsson P and Splettstoesser J 2019 {\em
		Phys. Rev. B\/} {\bf 100} 035405
	
	\bibitem{bisognin2019}
	Bisognin R, Marguerite A, Roussel B, Kumar M, Cabart C, Chapdelaine C,
	Mohammad-Djafari A, Berroir J~M, Bocquillon E, Pla\c{c}ais B, Cavanna A,
	Gennser U, Jin Y, Degiovanni P and F\`eve G 2019 {\em Nature
		Communications\/} {\bf 10} 3379 ISSN 2041-1723
	
	\bibitem{burset19}
	Burset P, Kotilahti J, Moskalets M and Flindt C 2019 {\em Advanced Quantum
		Technologies\/} {\bf 2} 1900014
	
	\bibitem{tersoff85}
	Tersoff J and Hamann D~R 1985 {\em Phys. Rev. B\/} {\bf 31} 805--813
	
	\bibitem{wiesendanger94}
	Wiesendanger R 1994 {\em Scanning Probe Microscopy and Spectroscopy: Methods
		and Applications\/} (Cambridge University Press)
	
	\bibitem{chen07}
	Chen C~J 2007 {\em Introduction to Scanning Tunneling Microscopy: Second
		Edition\/} (Oxford University Press)
	
	\bibitem{auslaender02}
	Auslaender O~M, Yacoby A, de~Picciotto R, Baldwin K~W, Pfeiffer L~N and West
	K~W 2002 {\em Science\/} {\bf 295} 825
	
	\bibitem{steinberg07chargefrac}
	Steinberg H, Barak G, Yacoby A, Pfeiffer L~N, West K~W, Halperin B~I and Le~Hur
	K 2007 {\em Nature Physics\/} {\bf 4} 117
	
	\bibitem{lehur08}
	Hur K~L, Halperin B~I and Yacoby A 2008 {\em Annals of Physics\/} {\bf 323}
	3037
	
	\bibitem{crepieux03}
	Cr\'epieux A, Guyon R, Devillard P and Martin T 2003 {\em Phys. Rev. B\/} {\bf
		67} 205408
	
	\bibitem{lebedev05}
	Lebedev A~V, Cr\'epieux A and Martin T 2005 {\em Phys. Rev. B\/} {\bf 71}
	075416
	
	\bibitem{guigou07}
	Guigou M, Popoff A, Martin T and Cr\'epieux A 2007 {\em Phys. Rev. B\/} {\bf
		76} 045104
	
	\bibitem{guigou09}
	Guigou M, Martin T and Cr\'epieux A 2009 {\em Phys. Rev. B\/} {\bf 80} 045420
	
	\bibitem{guigou09-2}
	Guigou M, Martin T and Cr\'epieux A 2009 {\em Phys. Rev. B\/} {\bf 80} 045421
	
	\bibitem{dasprl2011}
	Das S and Rao S 2011 {\em Phys. Rev. Lett.\/} {\bf 106} 236403
	
	\bibitem{liu15-tip3dti}
	Liu L, Richardella A, Garate I, Zhu Y, Samarth N and Chen C~T 2015 {\em Phys.
		Rev. B\/} {\bf 91} 235437
	
	\bibitem{hus17-tip3dti}
	Hus S~M, Zhang X~G, Nguyen G~D, Ko W, Baddorf A~P, Chen Y~P and Li A~P 2017
	{\em Phys. Rev. Lett.\/} {\bf 119} 137202
	
	\bibitem{voigtlander18-tip}
	Voigtl\"ander B, Cherepanov V, Korte S, Leis A, Cuma D, Just S and Lüpke F
	2018 {\em Review of Scientific Instruments\/} {\bf 89} 101101
	
	\bibitem{Giordano:18}
	Giordano M~C {\em et~al.\/} 2018 {\em Opt. Express\/} {\bf 26} 18423
	
	\bibitem{grenier11tomography}
	Grenier C, Herv\'e R, Bocquillon E, Parmentier F~D, Pla{\c c}ais B, Berroir
	J~M, F\`eve G and Degiovanni P 2011 {\em New Journal of Physics\/} {\bf 13}
	093007
	
	\bibitem{voit1995}
	Voit J 1995 {\em Reports on Progress in Physics\/} {\bf 58} 977
	
	\bibitem{stuhler19}
	{St{\"u}hler} R, {Reis} F, {M{\"u}ller} T, {Helbig} T, {Schwemmer} T, {Thomale}
	R, {Sch{\"a}fer} J and {Claessen} R 2019   arXiv:1901.06170
	
	\bibitem{li2015helical}
	Li T, Wang P, Fu H, Du L, Schreiber K~A, Mu X, Liu X, Sullivan G, Cs\'athy G~A,
	Lin X and Du R~R 2015 {\em Phys. Rev. Lett.\/} {\bf 115} 136804
	
	\bibitem{glazman16}
	V\"ayrynen J~I, Geissler F and Glazman L~I 2016 {\em Phys. Rev. B\/} {\bf 93}
	241301
	
	\bibitem{vondelft}
	von Delft J and Schoeller H 1998 {\em Ann. Phys.\/} {\bf 7} 225
	
	\bibitem{giamarchi}
	Giamarchi T 2003 {\em Quantum Physics in One Dimension\/} (Oxford University
	Press)
	
	\bibitem{safi95}
	Safi I and Schulz H~J 1995 {\em Phys. Rev. B\/} {\bf 52} R17040
	
	\bibitem{pham2000}
	Pham K~V, Gabay M and Lederer P 2000 {\em Phys. Rev. B\/} {\bf 61} 16397
	
	\bibitem{auslaender05}
	Auslaender O~M, Steinberg H, Yacoby A, Tserkovnyak Y, Halperin B~I, Baldwin
	K~W, Pfeiffer L~N and West K~W 2005 {\em Science\/} {\bf 308} 88
	
	\bibitem{jompol09}
	Jompol Y, Ford C~J~B, Griffiths J~P, Farrer I, Jones G~A~C, Anderson D, Ritchie
	D~A, Silk T~W and Schofield A~J 2009 {\em Science\/} {\bf 325} 597
	
	\bibitem{deshpande10}
	Deshpande V~V, Bockrath M, Glazman L~I and Yacoby A 2010 {\em Nature\/} {\bf
		464} 209
	
	\bibitem{perfetto14timeresolved}
	Perfetto E, Stefanucci G, Kamata H and Fujisawa T 2014 {\em Phys. Rev. B\/}
	{\bf 89} 201413
	
	\bibitem{guinea95}
	Guinea F, Santos G~G, Sassetti M and Ueda M 1995 {\em Europhysics Letters
		({EPL})\/} {\bf 30} 561
	
	\bibitem{calzona17quench}
	Calzona A, Gambetta F~M, Carrega M, Cavaliere F and Sassetti M 2017 {\em Phys.
		Rev. B\/} {\bf 95} 085101
	
	\bibitem{calzona17-2}
	Calzona A, Gambetta F~M, Cavaliere F, Carrega M and Sassetti M 2017 {\em Phys.
		Rev. B\/} {\bf 96} 085423
	
	\bibitem{gambetta:epl}
	Gambetta F~M {\em et~al.\/} 2014 {\em Europhys. Lett.\/} {\bf 107} 47010
	
	\bibitem{ferraro18squeezing}
	Ferraro D, Ronetti F, Rech J, Jonckheere T, Sassetti M and Martin T 2018 {\em
		Phys. Rev. B\/} {\bf 97} 155135
	
	\bibitem{ferraro14decoherence}
	Ferraro D, Roussel B, Cabart C, Thibierge E, F\`eve G, Grenier C and Degiovanni
	P 2014 {\em Phys. Rev. Lett.\/} {\bf 113} 166403
	
	\bibitem{acciai19}
	Acciai M, Ronetti F, Ferraro D, Rech J, Jonckheere T, Sassetti M and Martin T
	2019 {\em Phys. Rev. B\/} {\bf 100} 085418
	
	\bibitem{fisher_review}
	Fisher M~P~A and Glazman L 1997 {\em Mesoscopic Electron Transport (NATO ASI
		Series E: Applied Science)\/} {\bf 345} 331--374
	
	\bibitem{vanevic07}
	Vanevi\ifmmode~\acute{c}\else \'{c}\fi{} M, Nazarov Y~V and Belzig W 2007 {\em
		Phys. Rev. Lett.\/} {\bf 99} 076601
	
	\bibitem{belzig16physe}
	Belzig W and Vanevic M 2016 {\em Physica E: Low-dimensional Systems and
		Nanostructures\/} {\bf 75} 22
	
\end{thebibliography}

\providecommand{\newblock}{}

\end{document}